\documentclass[12pt]{article}
\usepackage{amsmath}
\usepackage{amsthm}
\usepackage{amssymb}
\usepackage{amsfonts}
\usepackage{epic}
\usepackage{eepic}
 \usepackage{times}
\usepackage[matrix,arrow]{xy}
\usepackage{macros}
\usepackage{epsfig}
\usepackage{graphicx}
\usepackage{bm}

\theoremstyle{plain}
\newtheorem{cor}{Corollary}
\newtheorem{thm}{Theorem}

\theoremstyle{remark}

\newcommand{\Li}{{\rm Li}}

\newcommand{\pa}{\partial}
\newcommand{\ot}{\otimes}
\newcommand{\ra}{\to}

\newcommand{\fr}[2]{{\textstyle \frac{#1}{#2} }}

\newcommand{\fsl}{{\mathfrak s}{\mathfrak l}}

\DeclareMathOperator*{\Res}{Res}
\newcommand{\al}{\alpha}

\newcommand{\ga}{\gamma}
\newcommand{\Ga}{\Gamma}
\newcommand{\de}{\delta}
\newcommand{\De}{\Delta}
\newcommand{\ep}{\epsilon}

\newcommand{\si}{\sigma}
\newcommand{\up}{\Upsilon}

\newcommand{\bz}{{\bar{z}}}

\newcommand{\CA}{{\mathcal A}}
\newcommand{\CB}{{\mathcal B}}

\newcommand{\CC}{{\mathcal C}}

\newcommand{\CE}{{\mathcal E}}
\newcommand{\CF}{{\mathcal F}}
\newcommand{\CG}{{\mathcal G}}

\newcommand{\CI}{{\mathcal I}}
\newcommand{\CJ}{{\mathcal J}}
\newcommand{\CK}{{\mathcal K}}
\newcommand{\CL}{{\mathcal L}}
\newcommand{\CM}{{\mathcal M}}
\newcommand{\CN}{{\mathcal N}}
\newcommand{\CO}{{\mathcal O}}
\newcommand{\CP}{{\mathcal P}}

\newcommand{\CS}{{\mathcal S}}
\newcommand{\CT}{{\mathcal T}}

\newcommand{\CU}{{\mathcal U}}

\def\rank{\operatorname{rank}}

\newcommand{\SC}{{\mathsf C}}

\newcommand{\SE}{{\mathsf E}}
\newcommand{\SF}{{\mathsf F}}

\newcommand{\SK}{{\mathsf K}}
\newcommand{\sk}{{\mathsf k}}
\newcommand{\SL}{{\mathsf L}}

\newcommand{\SR}{{\mathsf R}}

\newcommand{\sll}{{\mathsf l}}

\newcommand{\spp}{{\mathsf p}}

\newcommand{\sx}{{\mathsf x}}

\newcommand{\BR}{{\mathbb R}}

\newcommand{\BC}{{\mathbb C}}

\newcommand{\BZ}{{\mathbb Z}}

\newcommand{\rf}[1]{(\ref{#1})}
\newcommand{\fus}[6]{\big\{\,{}^{#1}_{#3}\;{}^{#2}_{#4}\;{}^{#5}_{#6}\,
\big\}_b} 

\newcommand{\nc}{\newcommand}

\nc{\rnc}{\renewcommand} \nc{\beq}{\begin{equation}}
\nc{\eeq}{\end{equation}} \nc{\beqa}{\begin{eqnarray}}
\nc{\eeqa}{\end{eqnarray}}

\def\stackreb#1#2{\ \mathrel{\mathop{#1}\limits_{#2}}}

\begin{document}
\title{6j symbols for the modular double, 
quantum hyperbolic geometry, 
and supersymmetric gauge theories}
\author{J. Teschner and G. S. Vartanov}
\address{
DESY Theory, Notkestr. 85, 22603 Hamburg, Germany}
\maketitle

\begin{quote}
\centerline{\bf Abstract}
{\small We revisit the definition of the $6j$ symbols from the 
modular double of $\CU_q(\fsl(2,\BR))$, referred to as b-$6j$ symbols. 
Our new results 
are (i) the identification of particularly natural normalization
conditions, and (ii) new integral representations for this
object. This is used to briefly discuss possible applications
to quantum hyperbolic geometry, and to the study of
certain supersymmetric gauge theories. We show, in particular,
that the b-$6j$ symbol has
leading semiclassical asymptotics given by
the volume of a non-ideal tetrahedron. We furthermore 
observe
a close relation with the problem to quantize
natural Darboux coordinates for moduli spaces
of flat connections on Riemann surfaces related to the 
Fenchel-Nielsen coordinates. 
Our new integral representations
finally indicate a possible interpretation of 
the b-$6j$ symbols as partition functions of non-abelian
three-dimensional $\mathcal{N}=2$ supersymmetric 
gauge theories.
}\end{quote}

\section{Introduction}
\setcounter{equation}{0}

Analogs of the Racah-Wigner $6j$-symbols coming from the study
of a non-compact quantum group have been introduced in \cite{PT1}.
The quantum group in question is related to $\CU_q(\fsl(2,\BR))$
and is often referred to as the modular double of 
$\CU_q(\fsl(2,\BR))$. The $6j$-symbols of this quantum 
group, which will be called b-$6j$ symbols, 
play an important role for the harmonic analysis of the 
modular double \cite{PT2},
quantum Liouville theory \cite{Teschner:2001rv}
and quantum Teichm\"uller theory \cite{Teschner:2003at}.
The terminology b-$6j$ symbol is partly motivated by the 
fact that it is useful 
to parameterize
the deformation parameter $q$ of $\CU_q(\fsl(2,\BR))$
in terms of a parameter $b$ as $q=e^{\pi \textup{i} b^2}$.

However, the precise definition of the b-$6j$
depends on the normalization of the Clebsch-Gordan maps. 
Similar normalization issues arise in 
Liouville theory and in quantum Teich\-m\"uller theory. In the case of 
Liouville theory it is related to the issue to fix normalizations 
for bases in the space of conformal blocks. In quantum 
Teichm\"uller theory it is related to the precise definition
of the representations in which a maximal commuting set of
geodesic length operators is diagonal.
The normalizations chosen in the references above were somewhat adhoc.
One of our first goals in this paper is to discuss
natural ways to fix this issue.

We will show that there exist
very natural normalizations which also appear to be 
very natural
from the point of view of Liouville- and the
quantum Teichm\"uller theory. In the latter context, one of the
normalizations defining our b-$6j$ symbols will be shown 
to define a quantization of the Fenchel-Nielsen coordinates.
Somewhat strikingly, we will 
find that the b-$6j$ symbols defined in this way exactly 
reproduce the
hyperbolic volume of a non-ideal tetrahedron with given
dihedral angles in the classical limit $b\ra 0$. 
This strongly suggests that Turaev-Viro type \cite{TuV}
state-sum models built from the b-$6j$ symbols are related
to three-dimensional quantum gravity with negative cosmological
constant, which can be seen as an analog
of earlier observations for the cases of 
zero \cite{PR} and positive cosmological 
constants \cite{MT}, respectively. The b-$6j$ symbols are 
also natural building blocks for combinatorial approaches 
to  the quantization of $SL(2,\BR)$-Chern-Simons theory or of its
complexification.

One of our main technical 
results will be new integral representations for
the b-$6j$ symbols. One of them strongly 
resembles the formulae for the usual $6j$ symbols.
The new integral representations will be obtained from the 
formula for the b-$6j$ symbols obtained in \cite{PT2} 
by a sequence of nontrivial integral transformations that 
follow from an identity satisfied by Spiridonov's 
elliptic hypergeometric integrals \cite{S1,S2} 
(for a review see \cite{S3}) in certain limits.
We will point out that one of these integral
representations admits an interpretation as 
a partition function for a non-abelian three-dimensional supersymmetric
gauge theory. This, and the relations to three-dimensional
Chern-Simons theories mentioned above suggest that
the b-$6j$ symbols could play a key role in the currently 
investigated program 
to identify correspondences between three-dimensional 
supersymmetric gauge theories and noncompact 
Chern-Simons theories on suitable three-manifolds 
\cite{Terashima:2011qi,DiGu,
Dimofte:2011ju}.


\section{Racah-Wigner 6j symbols for the modular double}
\setcounter{equation}{0}

\subsection{Self-dual representations of $\CU_q(\fsl(2,\BR))$ and the modular
double}

We will be considering the Hopf-algebra $\CU_q(\fsl(2,\BR))$
which has generators $E$, $F$ and $K$ subject to the usual 
relations. This algebra has a one-parameter
family of representations $\CP_{\al}$
\begin{equation}\label{Paldef}
\begin{aligned}
 \SE_\al
\equiv\pi_{\al}(E):=e^{+\pi b\sx}\frac{\cosh\pi b (\spp-s)}{\sin\pi b^2}
e^{+\pi b\sx}\,,\\
\SF_\al\equiv\pi_{\al}(F):=e^{-\pi b\sx}\frac{\cosh\pi b (\spp+s)}{\sin\pi b^2}
e^{-\pi b\sx}\,,
\end{aligned}
\qquad \SK_\al\equiv\pi_\al(K):=e^{-\pi b\spp}\,,
\end{equation}
where $\spp$ and $\sx$ are operators acting on functions $f(x)$ as
$\spp f(x)=(2\pi i)^{-1}\frac{\pa}{\pa x}f(x)$ and $\sx f(x)=xf(x)$,
respectively.
In the definitions \rf{Paldef} we are parameterizing $q$ as
$q=e^{\pi \textup{i} b^2}$, and write the parameter $\al$ as 
$\al=Q/2+\textup{i}s$. There is a maximal dense 
subspace $\CP_{\al}\subset L^2(\BR)$ on which all polynomials formed out of 
$\SE_\al$, $\SF_\al$ and $\SK_\al$ are well-defined 
\cite[Appendix B]{Bytsko:2006ut}.

These representations are distinguished by
a remarkable self-duality property: They are automatically 
representations of the quantum group 
$\CU_{\tilde{q}}(\fsl(2,\BR))$, where 
$\tilde{q}=e^{\pi \textup{i}/b^2}$ if $q=e^{\pi \textup{i} b^2}$. 
These representations are generated from operators 
$\tilde{\SE}_\al$, $\tilde{\SF}_\al$ and  $\tilde{\SK}_\al$ 
which are defined by formulae obtained from 
those in \rf{Paldef} by replacing $b\ra b^{-1}$.
The subspace $\CP_\al$ is simultaneously 
a maximal domain for the polynomial functions of 
$\tilde{\SE}_\al$, $\tilde{\SF}_\al$ and  $\tilde{\SK}_\al$ 
\cite[Appendix B]{Bytsko:2006ut}.

This phenomenon was observed independently in \cite{PT1} and in 
\cite{FMD}. 
It is closely related to the fact that  $\SE_\al$, $\SF_\al$ and $\SK_\al$
are  {\it positive} 
self-adjoint generators which allows one to construct
$\tilde{\SE}_\al$, $\tilde{\SF}_\al$ and  $\tilde{\SK}_\al$
as $\SE^{1/b^2}_\al$
$\SF^{1/b^2}_\al$, $\SK^{{1}/{b^2}}_\al$ \cite{BT1}. 

It was proposed in \cite{PT1,BT1} to construct a
noncompact quantum group which has as {\it complete} set 
of tempered representations the self-dual representations
$\CP_{\al}$. 
It's gradually becoming clear how to realize this suggestion
precisely. 
Relevant steps in this direction were taken in \cite{BT1} by 
defining co-product, R-operator and Haar-measure of such a 
quantum group. Further important progress in this direction was
recently made in \cite{Ip}. 
Following \cite{FMD}, we will in the following call 
this noncompact quantum 
group the modular double of $\CU_q(\fsl(2,\BR))$.

\subsection{Normalized Clebsch-Gordan coefficients for the modular double}

The Clebsch-Gordan maps $\SC^{\al_3}_{\al_2,\al_1}:
\CP_{\al_2}\ot\CP_{\al_1}\ra\CP_{\al_3}$ were constructed in \cite{PT2}.
The defining intertwining property is
\begin{equation}\label{inter}
\SC^{\al_3}_{\al_2,\al_1}\cdot(\pi_{\al_2}\ot\pi_{\al_1})(\De(X))=
\pi_{\al_3}\cdot\SC^{\al_3}_{\al_2,\al_1}\,.
\end{equation}
In \cite{PT2} it was found that the $\SC^{\al_3}_{\al_2,\al_1}$
can be represented as integral operators
of the form
\begin{equation}\label{CGdef}
(\SC^{\al_3}_{\al_2,\al_1}\psi)(x_3)\,=\,
\int_{\BR^2}dx_1dx_2\;
\big(\,{}^{\al_3}_{x_3}\,|\,{}^{\al_2}_{x_2}\,{}^{\al_1}_{x_1}\big)_b
\,\psi(x_2,x_1)\,,
\end{equation}
The intertwining property \rf{inter} will be satisfied 
if we take 
$\big(\,{}^{\al_3}_{x_3}\,|\,{}^{\al_2}_{x_2}\,{}^{\al_1}_{x_1}\big)=
\big(\,{}^{\al_3}_{x_3}\,|\,{}^{\al_2}_{x_2}\,
{}^{\al_1}_{x_1}\big)_b^{\rm an}$, with 
\begin{align} \label{b-3j}
\big(\,{}^{\al_3}_{x_3}\,|\,{}^{\al_2}_{x_2}\,{}^{\al_1}_{x_1}\big)_b^{\rm an} &
 = e^{-\pi \textup{i}(\Delta_{\alpha_3}-\Delta_{\alpha_1}-\Delta_{\alpha_2})/2} D_{-\frac{\textup{i}}{2} (\alpha_1+\alpha_2+\alpha_3-Q)}
\big(x_2-x_1-\textup{i} \fr{\alpha_3}{2}\big)
 \\ 
& \quad\times D_{-\frac{\textup{i}}{2}(Q+\alpha_2-\alpha_3-\alpha_1)}^{}
\big(x_2-x_3-\textup{i} \fr{\alpha_1}{2}\big)
D_{-\frac{\textup{i}}{2}(Q+\alpha_1-\alpha_3-\alpha_2)}^{}
\big(x_3-x_1-\textup{i} \fr{\alpha_2}{2}\big)\,. 
\notag\end{align}
In \rf{b-3j} we are using the notations 
$\Delta_{\alpha} = \alpha (Q-\alpha)$ with $Q=b+b^{-1}$ and
\begin{equation}
D_{\textup{i} \alpha}(x) = \frac{S_b(Q/2 - \textup{i} x + \alpha)}{S_b(Q/2 - \textup{i} x - \alpha)}.
\end{equation}
$S_b(x)$ is the so-called double Sine-function which is closely 
related to the functions called quantum dilogarithm in \cite{FK2}
hyperbolic gamma function in \cite{Ru},
and quantum exponential function in \cite{Wo}. 
Definition and relevant properties 
are recalled in Appendix \ref{Qdil}.

One should note, however, that our definition of 
the $3j$ coefficients \rf{b-3j} is not canonical,  we might equally
well use
$\big(\,{}^{\al_3}_{x_3}\,|\,{}^{\al_2}_{x_2}\,{}^{\al_1}_{x_1}\big)_b'$ 
in \rf{CGdef},  with
\begin{equation} \label{3jrenorm}
\big(\,{}^{\al_3}_{x_3}\,|\,{}^{\al_2}_{x_2}\,{}^{\al_1}_{x_1}\big)_{b}':=
M(\al_3,\al_2,\al_1)
\big(\,{}^{\al_3}_{x_3}\,|\,{}^{\al_2}_{x_2}\,
{}^{\al_1}_{x_1}\big)_b^{\rm an}\,.
\end{equation} 
This will satisfy \rf{inter} for arbitrary functions $M(\al_3,\al_2,\al_1)$.
A natural choice for $M(\al_3,\al_2,\al_1)$ can be determined by
requiring the Weyl-invariance of the Clebsch-Gordan maps.
In order to formulate this requirement, we will need the 
intertwining operator $\SR_{\al}:\CP_{\al}\ra\CP_{Q-\al}$
which can be represented explicitly as integral operator \cite{PT2}
\begin{equation}
(\SR_{\al}f)(x):=S_b(2\al)
\int_\BR dx'\;D_{-\textup{i}\al}(x-x')f(x)\,.
\end{equation}
We may now require that
\begin{equation}\label{Weyl}
\begin{aligned}
&
\SC^{\al_3}_{\al_2,\al_1}\cdot (1\ot\SR_{Q-\al_1})=
\SC^{\al_3}_{\al_2,Q-\al_1}\,,
\\
&\SC^{\al_3}_{\al_2,\al_1}\cdot (\SR_{Q-\al_2}\ot 1)
=\SC^{\al_3}_{Q-\al_2,\al_1}\,,
\end{aligned}\qquad
\SR_{\al_3}\cdot \SC^{\al_3}_{\al_2,\al_1}=\SC^{Q-\al_3}_{\al_2,\al_1}\,.
\end{equation}
We claim that \rf{Weyl} is satisfied if
we choose $M(\al_3,\al_2,\al_1)$ as 
\begin{align} \label{normcoef}
& M(\alpha_3,\alpha_2,\alpha_1)= \\ \nonumber & = 
\big(S_b(2Q-\alpha_1-\alpha_2-\alpha_3) S_b(Q-\alpha_1-\alpha_2+\alpha_3) 
S_b(\alpha_1+\alpha_3-\alpha_2) 
S_b(\alpha_2+\alpha_3-\alpha_1)\big)^{-\frac{1}{2}}.
\end{align}

To prove this claim let us consider, for example, 
the first of the equations in
\rf{Weyl}, which would follow from the identity 
\beq \label{intertw}
S_b(2\bar{\al}_1)
\int_{\BR}dx_1' \;
\big(\,{}^{\al_3}_{x_3}\,|\,{}^{\al_2}_{x_2}\,
{}^{\al_1}_{x_1'}\big)_b^{\rm an}
D_{-\textup{i} \bar{\al}_1}(x_1'-x_1)   =
\xi\, \big(\,{}^{\al_3}_{x_3}\,|\,{}^{\al_2}_{x_2}\,
{}^{\bar{\al}_1}_{x_1}\big)_b^{\rm an}\,,  
\eeq
where we use abbreviation $\bar{\alpha}=Q-\alpha$ and
$
\xi 
= S_b(\alpha_2+\al_3-\alpha_1) S_b(2Q-\alpha_1-\alpha_2-\alpha_3).
$
This identity can easily be rewritten in the form \cite[Equation (A.34)]{BT1}
in which it is recognized 
as the famous star-triangle relation, 
see e.g. \cite{BMS}. Proofs can be found in 
\cite{K2,V}.
It can also be derived easily
from the so-called elliptic beta-integral \cite{S1} 
following the strategy discussed in Appendix \ref{idproof}.

We will denote the Clebsch-Gordan coefficients
defined by  \rf{3jrenorm} with function $M(\al_3,\al_2,\al_1)$ 
given in \rf{normcoef} 
as $\big(\,{}^{\al_3}_{x_3}\,|\,{}^{\al_2}_{x_2}\,
{}^{\al_1}_{x_1}\big)_b^{}$. We would like to stress that 
both $\big(\,{}^{\al_3}_{x_3}\,|\,{}^{\al_2}_{x_2}\,
{}^{\al_1}_{x_1}\big)_b^{\rm an}$ and 
$\big(\,{}^{\al_3}_{x_3}\,|\,{}^{\al_2}_{x_2}\,
{}^{\al_1}_{x_1}\big)_b^{}$ have their virtues. 
While $\big(\,{}^{\al_3}_{x_3}\,|\,{}^{\al_2}_{x_2}\,
{}^{\al_1}_{x_1}\big)_b^{}$ has more natural symmetry
properties, the virtue of 
$\big(\,{}^{\al_3}_{x_3}\,|\,{}^{\al_2}_{x_2}\,
{}^{\al_1}_{x_1}\big)_b^{\rm an}$ is to have nice
analytic properties in all of its variables.

\subsection{Normalized b-6j symbols for the modular double}

The composition of Clebsch-Gordan maps allows us to define two natural
families of projection operators 
\begin{align}
&({}_{}^s\SC^{\al_4}_{\al_3,\al_2,\al_1}(\al_s)\Psi)(x_4)\,=\,
\int_{\BR^3}dx_1dx_2dx_3\;
\CE_{\al_s}^{(s)}(A|X)\psi(x_3,x_2,x_1)\,,\\
&({}_{}^t\SC^{\al_4}_{\al_3,\al_2,\al_1}(\al_t)\Psi)(x_4)\,=\,
\int_{\BR^3}dx_1dx_2dx_3\;
\CE_{\al_t}^{(t)}(A|X)\psi(x_3,x_2,x_1)\,,
\end{align}
with integral kernels $\CE_{\al_s}^{(s)}(A|X)$
and $\CE_{\al_t}^{(t)}(A|X)$ given as
\begin{align}
&\CE_{\al_s}^{(s)}(A|X)\,=\,\int dx_s\;
\big(\,{}^{\al_4}_{x_4}\,|\,{}^{\al_3}_{x_3}\,{}^{\al_s}_{x_s}\big)_b
\big(\,{}^{\al_s}_{x_s}\,|\,{}^{\al_2}_{x_2}\,{}^{\al_1}_{x_1}\big)_b\,,\\
&\CE_{\al_t}^{(t)}(A|X)\,=\,\int dx_t\;
\big(\,{}^{\al_4}_{x_4}\,|\,{}^{\al_t}_{x_t}\,{}^{\al_1}_{x_1}\big)_b
\big(\,{}^{\al_t}_{x_t}\,|\,{}^{\al_3}_{x_3}\,{}^{\al_2}_{x_2}\big)_b\,.
\end{align}

The b-$6j$ symbols $\big\{\,{}^{\al_1}_{\al_3}\,{}^{\al_2}_{{\al}_4}\,|\,{}^{\al_s}_{\al_t}\big\}_b$ are then defined by the relations
\begin{equation}\label{fusion'}
\CE_{\al_s}^{(s)}(A|Z)\,=\,\int_{{Q/2}+\textup{i}\BR} d\mu(\al_t)\;
\big\{\,{}^{\al_1}_{\al_3}\,{}^{\al_2}_{{\al}_4}\,|\,{}^{\al_s}_{\al_t}\big\}_b\,
\CE_{\al_t}^{(t)}(A|Z)\,,
\end{equation}
where the Plancherel measure $d\mu(\al)$ is explicitly given by
the expression
\begin{equation} \label{Pmeas}
d\mu(\al)\,=\,d\al\;M(\al)\,,\qquad M(\al):=|S_b(2\al)|^2\,.
\end{equation}
It is clear that the explicit expression for the b-$6j$ symbols
depends on the normalization chosen for the
Clebsch-Gordan maps. We will denote 
the $6j$ symbols corresponding to 
$\big(\,{}^{\al_3}_{x_3}\,|\,{}^{\al_2}_{x_2}\,
{}^{\al_1}_{x_1}\big)_b^{\rm an}$
and 
$\big(\,{}^{\al_3}_{x_3}\,|\,{}^{\al_2}_{x_2}\,
{}^{\al_1}_{x_1}\big)_b^{}$ by
$\big\{\,{}^{\al_1}_{\al_3}\,{}^{\al_2}_{{\al}_4}\,|\,{}^{\al_s}_{\al_t}\big\}_b^{\rm an}$
and
$\big\{\,{}^{\al_1}_{\al_3}\,{}^{\al_2}_{{\al}_4}\,|\,{}^{\al_s}_{\al_t}\big\}_b^{}$,
respectively.

The b-$6j$ symbols $\big\{\,{}^{\al_1}_{\al_3}\,{}^{\al_2}_{{\al}_4}\,|\,{}^{\al_s}_{\al_t}\big\}_b^{\rm an}$ were calculated in \cite{PT2}\footnote{The formula below coincides with equation (228) in \cite{Teschner:2001rv} after shifting $s \rightarrow u-\alpha_s-Q/2$. We have moved a factor $|S_b(2\al_t)|^2$ into the
measure of integration in \rf{fusion'}.}, 
\begin{align} \label{6j1}
\big\{\,{}^{\al_1}_{\al_3}\,{}^{\al_2}_{\bar{\al}_4}\,|\,{}^{\al_s}_{\al_t}\big\}_b^{\rm an}= & 
\;\frac{S_b(\alpha_2+\alpha_s-\alpha_1) S_b(\alpha_t+\alpha_1-\alpha_4)}{S_b(\alpha_2+\alpha_t-\alpha_3) S_b(\alpha_s+\alpha_3-\alpha_4)} \\
& \times  \int_{\CC}du\; 
S_b(-\alpha_2 \pm (\alpha_1-Q/2)+u) 
S_b(-\alpha_4 \pm (\alpha_3-Q/2) +u) \nonumber \\[-.5ex] &
\hspace{1.125cm}\times  
S_b(\alpha_2 + \alpha_4 \pm (\alpha_t-Q/2) - u) 
S_b(Q \pm (\alpha_s - Q/2) - u) \,.
\notag\end{align}
The following notation has been used
$
S_b(\alpha \pm u) :=
S_b(\alpha+u)
S_b(\alpha-u).
$
The integral in \rf{6j1} will be defined for $\al_k\in Q/2+\textup{i}\BR$ by
using a contour $\CC$ that approaches $Q+\textup{i}\BR$ near infinity,
and passes the real axis in $(Q/2,Q)$, and for other values 
of $\al_k$ by analytic continuation.

The b-$6j$ symbols corresponding to the 
normalization defined above are then given by the formula
\begin{equation} \label{norm}
\big\{\,{}^{\al_1}_{\al_3}\,{}^{\al_2}_{{\al}_4}\,|\,{}^{\al_s}_{\al_t}\big\}_b^{}={\textstyle
\frac{M(\alpha_s,\alpha_2,\alpha_1)M(\alpha_4,\alpha_3,\alpha_s)}{M(\alpha_t,\alpha_3,\alpha_2) M(\alpha_4,\alpha_t,\alpha_1)} }
\big\{\,{}^{\al_1}_{\al_3}\,{}^{\al_2}_{{\al}_4}\,|\,{}^{\al_s}_{\al_t}\big\}_b^{\rm an}\,.
\end{equation}
with $M(\al_3,\al_2,\al_1)$ being defined in \rf{normcoef}.

\subsection{3j symbols for the modular double}

$3j$ coefficients describe invariants in tensor products of three 
representations. Such invariants may be constructed from 
the Clebsch-Gordon maps and the invariant bilinear form
$\CB:\CP_{\al}\ot\CP_{Q-\al}\ra \BC$ defined by \cite{PT2}
\begin{equation}
\CB(f,g):=\,\int_\BR dx\; f(x)g(x-\textup{i}Q/2)\,.
\end{equation}
We may thereby construct an invariant trilinear form 
$\CC_{\al_3,\al_2,\al_1}:\CP_{\al_3}\ot\CP_{\al_2}\ot\CP_{\al_1}\ra \BC$ as
\begin{equation}
\CC_{\al_3,\al_2,\al_1}(f_3,f_2,f_1):=\CB\big(\,f_3\,,\,
\SC^{Q-\al_3}_{\al_2,\al_1}\cdot f_2\ot f_1\,\big)\,.
\end{equation}
The form $\CC_{\al_3,\al_2,\al_1}$ can be represented as
\begin{equation}
\CC_{\al_3,\al_2,\al_1}(f_3,f_2,f_1)\,=\,
\int_{\BR^3}dx_3dx_2dx_1\;
\big(\,{}^{\al_3}_{x_3}\;{}^{\al_2}_{x_2}\;{}^{\al_1}_{x_1}\big)\,
f_3(x_3)f_2(x_2)f_1(x_1)
\end{equation}
with $3j$-symbols $\big(\,{}^{\al_3}_{x_3}\,{}^{\al_2}_{x_2}\,
{}^{\al_1}_{x_1}\big)$
given in terms of the Clebsch-Gordan coefficients
$\big(\,{}^{\al_3}_{x_3}\,|\,{}^{\al_2}_{x_2}\,{}^{\al_1}_{x_1}\big)$
as
\begin{equation}
\big(\,{}^{\al_3}_{x_3}\;{}^{\al_2}_{x_2}\;{}^{\al_1}_{x_1}\big)=
\big(\,{}^{Q-\al_3}_{x_3-\textup{i}Q/2}\,|\,{}^{\al_2}_{x_2}\,{}^{\al_1}_{x_1}\big)\,.
\end{equation}

We may similarly define 
\begin{equation}
\begin{aligned}
{}_{}^s\CC_{\al_4,\al_3,\al_2,\al_1}^{\al_s}(f_4,f_3,f_2,f_1):=
\CB\big(\,f_4\,,\,
{}_{}^s\SC^{Q-\al_4}_{\al_3,\al_2,\al_1}(\al_s)\cdot 
f_3\ot f_2\ot f_1\,\big)\,,\\
{}_{}^t\CC_{\al_4,\al_3,\al_2,\al_1}^{\al_t}(f_4,f_3,f_2,f_1):=
\CB\big(\,f_4\,,\,
{}_{}^t\SC^{Q-\al_4}_{\al_3,\al_2,\al_1}(\al_t)\cdot 
f_3\ot f_2\ot f_1\,\big)\,.
\end{aligned}
\end{equation}
The b-$6j$ symbols $\big\{\,{}^{\al_1}_{\al_3}\;{}^{\al_2}_{\al_4}\;{}^{\al_s}_{\al_t}\big\}$ are then defined by the relation
\begin{equation}\label{6j-def}
{}_{}^s\CC_{\al_4,\al_3,\al_2,\al_1}^{\al_s}
\,=\,\int_{{Q/2}+\textup{i}\BR} d\mu(\al_t)\;
\big\{\,{}^{\al_1}_{\al_3}\;{}^{\al_2}_{\al_4}\;{}^{\al_s}_{\al_t}\big\}_b\,
{}_{}^t\CC_{\al_4,\al_3,\al_2,\al_1}^{\al_t}\,.
\end{equation}
It follows that
\begin{equation}\label{6j-moddef}
\big\{\,{}^{\al_1}_{\al_3}\;{}^{\al_2}_{\al_4}\;{}^{\al_s}_{\al_t}\big\}_b
=
\big\{\,{}^{\al_1}_{\al_3}\,{}^{\al_2}_{\bar\al_4}\,|\,{}^{\al_s}_{\al_t}\big\}_b\,,
\qquad
\bar\al_4:=Q-\al_4.
\end{equation}
The b-$6j$ symbols satisfy
the following identities \cite{PT1}
\begin{equation}\begin{aligned}
& \int_{Q/2+\textup{i}\BR^+}d\mu(\delta_{1})\; 
\fus{\al_1}{\al_2}{\al_3}{\beta_{2}}{\beta_{1}}{\delta_{1}}
\fus{\al_1}{\delta_1}{\al_4}{\al_5}{\beta_{2}}{\gamma_{2}}
\fus{\al_2}{\al_3}{\al_4}{\gamma_{2}}{\delta_{1}}{\gamma_{1}} 
=\fus{\beta_1}{\al_3}{\al_4}{\al_5}{\beta_{2}}{\gamma_{1}}
\fus{\al_1}{\al_2}{\gamma_1}{\al_{5}}{\beta_{1}}{\gamma_{2}}\,,\\
& \int_{Q/2+\textup{i}\BR^+}d\mu(\al_s)\;
\fus{\al_1}{\al_2}{\al_3}{\al_4}{\al_s}{\al_t}^*\,
\fus{\al_1}{\al_2}{\al_3}{\al_4}{\al_s}{\al_t'}
\,=\,(M(\al_t))^{-1}\de(\al_t-\al_t')\,.
\end{aligned}
\end{equation}
The explicit expression will again depend on the chosen normalization
of the Clebsch-Gordan maps, giving us two versions,
$\big\{\,{}^{\al_1}_{\al_3}\;{}^{\al_2}_{\al_4}\;{}^{\al_s}_{\al_t}\big\}_b^{}$
and
$\big\{\,{}^{\al_1}_{\al_3}\;{}^{\al_2}_{\al_4}\;{}^{\al_s}_{\al_t}\big\}_b^{\rm an}$,
respectively.


\subsection{A new integral formula for the b-6j symbols}

One of our main results will be the following formula 
for the b-$6j$ symbols:
\begin{align}\label{6j3}
\big\{\,{}^{\al_1}_{\al_3}\;{}^{\al_2}_{\al_4}\;{}^{\al_s}_{\al_t}\big\}_b^{}
&=\Delta(\al_s,\al_2,\al_1)\Delta(\al_4,\al_3,\al_s)\Delta(\al_t,\al_3,\al_2)
\Delta(\al_4,\al_t,\al_1)\\
&\qquad \times\int\limits_{\CC}du\;
S_b(u-\alpha_{12s}) S_b(u-\al_{s34}) S_b(u -\alpha_{23t}) 
S_b(u-\alpha_{1t4})
\notag \\[-1.5ex] & \hspace{2cm}\times 
S_b( \alpha_{1234}-u) S_b(\alpha_{st13}-u) 
S_b(\alpha_{st24}-u) S_b(2Q-u)\,.
\notag\end{align}
The expression involves the following ingredients:
\begin{itemize}
\item We have used the notations $\al_{ijk}=\al_i+\al_j+\al_k$,
$\al_{ijkl}=\al_i+\al_j+\al_k+\al_l$. 
\item $\Delta(\al_3,\al_2,\al_1)$ is defined as 
\begin{align}
&\Delta(\al_3,\al_2,\al_1)=\bigg(\frac{S_b(\alpha_1+\alpha_2+\alpha_s-Q)}{S_b(\alpha_1+\alpha_2-\alpha_s) 
S_b(\alpha_1+\alpha_s-\alpha_2) S_b(\alpha_2+\alpha_s-\alpha_1)} 
\bigg)^{\frac{1}{2}}\,.
\notag\end{align}
\item The integral is defined in the cases
that $\al_k\in Q/2+\textup{i}\BR$ by a contour $\CC$ which 
approaches $2Q+\textup{i}\BR$ near infinity,
and passes the real axis in the interval $(3Q/2,2Q)$.
For other values of the variables $\al_k$ it is defined by analytic 
continuation.
\end{itemize}
The reader may notice how closely the structure of the expression 
in \rf{6j3} resembles the well-known formulae for the classical 
$6j$ symbols. 

For establishing this relation, the
main step is contained in the following integral identity:
\begin{align}\label{6j2}
\big\{\,{}^{\al_1}_{\al_3}\,{}^{\al_2}_{\bar\al_4}\,|\,
{}^{\al_s}_{\al_t}\big\}_b^{\rm an}=\mathcal{C}(\underline{\alpha}) 
\int_{\CC'} du\;& 
S_b(u-\alpha_{12s}) S_b(u-\al_{s34}) S_b(u -\alpha_{23t}) 
S_b(u-\alpha_{1t4})
\\ \times & 
S_b( \alpha_{1234}-u) S_b(\alpha_{st13}-u) 
S_b(\alpha_{st24}-u) S_b(2Q-u),
\notag\end{align}
where the contour $\CC'$ in \rf{6j2} runs between $2Q-\textup{i}\infty$
and $2Q+\textup{i}\infty$,
and $\underline{\alpha}$ is shorthand notation for the tuple 
$(\al_1,\al_2,\al_3,\al_4,\al_s,\al_t)$. The prefactor 
$\CC(\underline{\alpha})$ is explicitly given by the expression
\begin{align}
\mathcal{C}(\underline{\alpha}) =& 
\,S_b(-Q+\alpha_1+\alpha_4+\alpha_t) S_b(Q-\alpha_1-\alpha_2+\alpha_s) \nonumber \\ \times & \,S_b(-Q+\alpha_2+\alpha_3+\alpha_t) S_b(Q-\alpha_2+\alpha_3-\alpha_t) S_b(Q+\alpha_2-\alpha_3-\alpha_t) \\ 
\times & \,S_b(Q-\alpha_3+\alpha_4-\alpha_s) S_b(Q-\alpha_3-\alpha_4+\alpha_s) S_b(Q+\alpha_3-\alpha_4-\alpha_s). 
\notag\end{align}
The proof of identity \rf{6j2} is nontrivial. It is described in 
Appendix \ref{idproof}, based on recent advances in the theory
of elliptic generalizations of the 
hypergeometric functions \cite{S1,S2,S3}.

\section{Relations to three-dimensional hyperbolic geometry}
\setcounter{equation}{0}

Our goal in this section is to demonstrate by direct calculation 
that the b-$6j$ symbols
reproduce the volume of non-ideal tetrahedra in the classical limit.
A second, perhaps more
conceptual proof of this fact will be outlined in section 
\ref{sec:2hyp}
below. 

Similar observations concerning relations between the semiclassical
behavior of the noncompact quantum dilogarithm and hyperbolic 
volumes have previously been made in \cite{Hikami1,Hikami2,Hikami3,BMS,BMS2,DGLZ,AK}. 
It would be interesting to understand the precise relations to our
result below.

\subsection{Volumes of non-ideal tetrahedra}

We are considering non-ideal tetrahedra which are
completely defined by the collection of six dihedral 
angles $\eta_1,\dots,\eta_6$. In order to formulate the
formula for their volumes from \cite{M}, let us
use the notation $A_k=e^{\textup{i}\eta_k}$, and define
\begin{align} \label{Udef}
 U({u},\underline{{A}}) = & 
\Li_2({u}) +  \Li_2({A}_{st13} {u}) + \
\Li_2({A}_{st24} {u}) + \Li_2({A}_{1234} {u}) \\ & 
 - \Li_2(-{A}_{12s}{u}) - \Li_2(-{A}_{s34} {u}) - 
\Li_2(-{A}_{4t1} {u}) - \Li_2(-{A}_{32t} {u})\,,
\nonumber\end{align}
where $A_{ijk}:=A_iA_jA_k$, $A_{ijkl}:=A_iA_jA_kA_l$, along with
\begin{align}\label{Deltadef}
 \Delta(\underline{{A}}) &= \log {A}_s {A}_t + \log {A}_2 {A}_4 + \log {A}_1 {A}_3  \\ & \quad + \tilde{\Delta}({A}_s,{A}_1,{A}_2) + \tilde{\Delta}({A}_s,{A}_3,{A}_3) + \tilde{\Delta}({A}_t,{A}_1,{A}_4) + \tilde{\Delta}({A}_t,{A}_2,{A}_3)\,, 
\notag\end{align}
where
\begin{align}
 \tilde{\Delta}({A}_1,{A}_2,{A}_3) = -\frac12 \big( & \Li_2(-{A}_1{A}_2{A}_3^{-1}) + \Li_2(-{A}_1{A}_2^{-1}{A}_3) + \Li_2(-{A}_1^{-1}{A}_2{A}_3) \nonumber \\ & + \Li_2(-{A}_1^{-1}{A}_2^{-1}{A}_3^{-1}) + \log^2 {A}_1 + \log^2 {A}_2 + \log^2 {A}_3 \big).
\notag\end{align}
The following formula was found in 
\cite[Theorem 2]{M} 
\beq \label{volume}
{\rm Vol}(\underline{{A}}) = \frac 12 \,{\rm Im}\big[ 
U(u_+,\underline{{A}})+ \Delta(\underline{{A}})\big] 
= - \frac 12  \,{\rm Im}\big[U(u_-,\underline{{A}})+ 
\Delta(\underline{{A}})\big]\,,
\eeq
where $u_\pm$ are the two roots of the equation
\beq \label{eq}
\frac{dU(u,\underline{{A}})}{du} \ = \ - \frac{2 \pi \textup{i}}{u}\,.
\eeq
It can be shown \cite{M} that equation (\ref{eq}) is a 
quadratic equation which has two 
solutions $u_\pm$ which are 
pure phase, $|u_{\pm}| =  1$.

\subsection{Semiclassical limit}

In the
following we will assume that $\al_k\in\BR$,
$0<\al_k<Q/2$.
In order to 
study the quasi-classical limit of (\ref{6j3}) let us write the 
right hand side
 of \rf{6j3} in the form 
\begin{equation}\label{Int}
I:=E(\underline{\al})\int_{\CC}du \,\CI(a,b;u)\,.
\end{equation}
The integrand $\CI(a,b;u)$ may be written as
\begin{equation}\label{int}
\CI(a,b;u)=
\frac{\prod_{i=1}^4 S_b(a_i+u)}{S_b(-Q+u) \prod_{i=1}^3 S_b(Q-b_i+u)} du,
\end{equation}
where
\begin{align}
& a\equiv[ a_1,a_2,a_3,a_4 ] \ = \ [-\alpha_s-\alpha_1-\alpha_2, -\alpha_s-\alpha_3-\alpha_4, -\alpha_t-\alpha_1-\alpha_4, -\alpha_t-\alpha_2-\alpha_3], \nonumber \\ & 
b\equiv[ b_1,b_2,b_3 ] \ = \ [ \alpha_s+\alpha_t+\alpha_1+\alpha_3, \alpha_s+\alpha_t+\alpha_2+\alpha_4, \alpha_1+\alpha_2+\alpha_3+\alpha_4 ].
\end{align}
The quasi-classical limit of $\CI(a,b;u)$ is easily determined with the help
of formula \rf{Sbscl} in Appendix \ref{Qdil}. In order to write
the result in an convenient form
let us reparameterize variables
$$e^{-2 \pi \textup{i} b  \alpha_k+\pi \textup{i}} \equiv A_k\,,\qquad k\in\{1,2,3,4,s,t\}\,.
$$
Introducing the integration variables $v:=2\pi b(u-Q/2)$ 
we get an integral of the form 
\begin{equation}\label{Int'}
I=D(\underline{\al})\int_{\CC-Q/2}
\frac{dv}{2\pi b} 
\,\CJ(a,b;v)
\end{equation} 
whose integrand $\CJ(a,b;v)$ 
has quasi-classical asymptotics
\begin{equation}
\CJ(a,b;v)\,=\,\exp\bigg(\frac{1}{2\pi b^2}\,U(e^{\textup{i}v},\underline{{A}})\bigg)
\Big(1+\CO(b^2)\Big)\,,
\end{equation}
with $U(e^{\textup{i}v},\underline{{A}})$ given by the formula \rf{Udef}.
The quasiclassical asymptotics of the prefactor in \rf{Int'} is
\beqa
&& D_{\rm cl}(\underline{{A}}) = \exp \left( \frac{1}{2 \pi \textup{i} b^2} 
\Big( \Delta(\underline{{A}}) - \frac53 \pi^2 \Big) \right)\,,
\eeqa
where $\Delta(\underline{{A}})$ was defined in \rf{Deltadef} above.

Now we are ready to perform the saddle-point approximation for 
the integral (\ref{Int}). 
The saddle points are the solutions of the equation \rf{eq}.
The values of the b-$6j$ at these points are
$$
\exp \bigg( \frac{1}{2 \pi \textup{i} b^2} 
W_{\pm}(\underline{{A}}) \bigg),\quad{\rm where}
\quad
W_{\pm}(\underline{{A}}) = U(z_{\pm},\underline{{A}}) + 
\Delta(\underline{{A}}) - \frac53 \pi^2 + 2 \pi \textup{i} \log u_{\pm}.
$$
Since $u_{\pm}=e^{\pm \pi \textup{i} \phi}$, $\phi\in\BR$ as noted
above,  we find that
\beq \label{quasclas}
W_{\pm}(\underline{{A}}) = U(z_{\pm},\underline{{A}}) + 
\Delta(\underline{{A}}) - \frac53 \pi^2 \mp 2 \pi^2 \phi.
\eeq
Taking the imaginary part of (\ref{quasclas}) one sees 
that we are  getting the volume 
of a hyperbolic tetrahedron \rf{volume}.

\section{Relation to Liouville theory and the representation 
theory of ${\rm Diff(S^1)}$}\label{Liou}

\setcounter{equation}{0}

In this section we want to explain that the normalization
leading to the definition of the b-$6j$ symbols is also 
very natural
from the point of view of Liouville theory.
This is closely related to the interpretation of b-$6j$ symbols
as $6j$ symbols for the the infinite-dimensional group
${\rm Diff(S^1)}$.

\subsection{Fusion kernel}

Recall that the fusion kernel is usually defined in terms 
of the conformal blocks appearing in the holomorphically 
factorized form of the four-point functions,
\begin{align}\notag
& \left\langle\, V_{\al_4}(z_4,\bz_4) V_{\al_3}(z_3,\bz_3)
V_{\al_2}(z_2,\bz_2)V_{\al_1}(z_1,\bz_1)\,\right\rangle
=\\ \label{s-channel}
&\qquad
=\int_{{Q}/{2}+\textup{i}\BR}d\al_s\;C(\al_4,\al_3,\al_s)C(Q-\al_s,\al_2,\al_1)
\CF_{\al_s}^{(s)}(A|Z)\CF_{\al_s}^{(s)}(A|\bar{Z})\,\\
&\qquad \label{t-channel}
=\int_{{Q}/{2}+\textup{i}\BR}d\al_t\;C(\al_4,\al_t,\al_1)C(Q-\al_t,\al_3,\al_2)
\CF_{\al_t}^{(t)}(A|Z)\CF_{\al_t}^{(t)}(A|\bar{Z})\,
\end{align}
where $A=(\al_1,\al_2,\al_3,\al_4)$, $Z=(z_1,z_2,z_3,z_4)$, and
\begin{align}\label{DOZZ}
C(\al_1 &,\al_2,\al_3)  =
(\pi \mu\ga(b^2)b^{2-2b^2})^{\frac{1}{b}(Q-\al_1-\al_2-\al_3)}\times \\
&\times\frac{\up_0\up(2\al_1)\up(2\al_2)\up(2\al_3)}
{\up(\al_1+\al_2+\al_3-Q)\up(\al_1+\al_3-\al_2)
\up(\al_1+\al_2-\al_3)\up(\al_2+\al_3-\al_1)
}
\,,
\notag\end{align}
here $\mu$ is the so-called cosmological constant in Liouville field 
theory and $\gamma(x)=\Ga(x)/\Ga(1-x)$. We also used $\up(x) = 
(\Ga_b(x)\Ga_b(Q-x))^{-1}$, $\up_0 = \frac{d \up(x)}{dx}|_{x=0}$ where 
the function $\Ga_b(x)$ is the Barnes double Gamma function.
Appendix A lists the definition and the relevant properties of $\Ga_b(x)$.

The first expression \rf{s-channel} for the four-point functions 
represents the operator product expansion of the fields 
$V_{\al_2}(z_2,\bz_2)$ and $V_{\al_1}(z_1,\bz_1)$, while the 
second expression \rf{t-channel} 
represents the operator product expansion of the fields 
$V_{\al_3}(z_3,\bz_3)$ and $V_{\al_2}(z_2,\bz_2)$. The equality of 
the two expressions \rf{s-channel} and \rf{t-channel} follows from the
validity of the relations 
\begin{equation}\label{fusion}
\CF_{\al_s}^{(s)}(A|Z)\,=\,\int_{{Q/2}+\textup{i}\BR} d\al_t\;F_{\al_s\al_t}
\big[\,{}^{\al_3}_{\al_4}\,{}^{\al_2}_{\al_1}\,\big]\,
\CF_{\al_t}^{(t)}(A|Z)\,,
\end{equation}
which were established in \cite{Teschner:2001rv}. The following formula was
found in \cite{PT1,Teschner:2001rv},
\begin{equation}\label{FvsRW}
F_{\al_s\al_t}
\big[\,{}^{\al_3}_{\al_4}\,{}^{\al_2}_{\al_1}\,\big]\,=\,
\frac{N(\alpha_s,\alpha_2,\alpha_1)N(\alpha_4,\alpha_3,\alpha_s)}
{N(\alpha_t,\alpha_3,\alpha_2) N(\alpha_4,\alpha_t,\alpha_1)} 
\,M(\al_t)\,\big\{\,{}^{\al_1}_{\al_3}\;{}^{\al_2}_{{\al}_4}\,|\,
{}^{\al_s}_{\al_t}\big\}^{\rm an}\,,
\end{equation}
where
\begin{align} \label{normcoefN}
& N(\alpha_3,\alpha_2,\alpha_1)= \\ \nonumber & = 
\frac{\Ga_b(2Q-2\al_3)\Ga_b(2\al_2)\Ga_b(2\al_1)}
{\Ga_b(2Q-\alpha_1-\alpha_2-\alpha_3) \Ga_b(Q-\alpha_1-\alpha_2+\alpha_3) 
\Ga_b(\alpha_1+\alpha_3-\alpha_2) \Ga_b(\alpha_2+\alpha_3-\alpha_1)}.
\end{align}

\subsection{Unitary normalization}

The expressions \rf{s-channel} and \rf{t-channel} strongly suggest to 
redefine the conformal blocks by absorbing the three-point functions
$C(\al_3,\al_2,\al_1)$ into the definition, 
\begin{equation}\label{Blnorm}\begin{aligned}
\CG_{\al_s}^{(s)}(A|Z):= 
\big(C(\al_4,\al_3,\al_s)C(Q-\al_s,\al_2,\al_1)\big)^{\frac{1}{2}}
\CF_{\al_s}^{(s)}(A|Z)\,,\\
\CG_{\al_t}^{(t)}(A|Z):= 
\big(C(\al_4,\al_t,\al_1)C(Q-\al_t,\al_3,\al_2)\big)^{\frac{1}{2}}
\CF_{\al_t}^{(t)}(A|Z)\,.
\end{aligned}
\end{equation}
This corresponds to normalizing the conformal blocks associated to the
three-punctured sphere in such a way that their scalar product is
always unity. This normalization may be called the unitary
normalization. We then have
\begin{align}
& \left\langle\, V_{\al_4}(z_4,\bz_4) V_{\al_3}(z_3,\bz_3)
V_{\al_2}(z_2,\bz_2)V_{\al_1}(z_1,\bz_1)\,\right\rangle
=\\ 
&\qquad
=\int_{\frac{Q}{2}+\textup{i}\BR}d\al_s\;
\CG_{\al_s}^{(s)}(A|Z)\CG_{\al_s}^{(s)}(A|\bar{Z})
=\int_{\frac{Q}{2}+\textup{i}\BR}d\al_t\;
\CG_{\al_t}^{(t)}(A|Z)\CG_{\al_t}^{(t)}(A|\bar{Z})\,,
\notag\end{align}
the second equation being a consequence of the unitarity of the 
change of basis
\begin{equation}\label{fusionL}
\CG_{\al_s}^{(s)}(A|Z)\,=\,\int_{{Q/2}+\textup{i}\BR} d\al_t\;
G_{\al_s\al_t}
\big[\,{}^{\al_3}_{\al_4}\,{}^{\al_2}_{\al_1}\,\big]\,
\CG_{\al_t}^{(t)}(A|Z)\,.
\end{equation}
The normalized fusion coefficients $G_{\al_s\al_t}
\big[\,{}^{\al_3}_{\al_4}\,{}^{\al_2}_{\al_1}\,\big]$ 
are related to the $F_{\al_s\al_t}
\big[\,{}^{\al_3}_{\al_4}\,{}^{\al_2}_{\al_1}\,\big]$ as
\begin{equation}\label{GvsF}
G_{\al_s\al_t}
\big[\,{}^{\al_3}_{\al_4}\,{}^{\al_2}_{\al_1}\,\big]\,=\,
\textstyle{ \sqrt{
\frac{C(\al_4,\al_3,\al_s)C(Q-\al_s,\al_2,\al_1)}
{C(\al_4,\al_t,\al_1)C(Q-\al_t,\al_3,\al_2)}}}\,
F_{\al_s\al_t}
\big[\,{}^{\al_3}_{\al_4}\,{}^{\al_2}_{\al_1}\,\big]\,.
\end{equation}
The fusion coefficients $G_{\al_s\al_t}
\big[\,{}^{\al_3}_{\al_4}\,{}^{\al_2}_{\al_1}\,\big]$ 
have a simple expression in terms of the b-$6j$ symbols,
\begin{equation} \label{eq2}
G_{\al_s\al_t}
\big[\,{}^{\al_3}_{\al_4}\,{}^{\al_2}_{\al_1}\,\big]\,=\,
\left({M(\al_t)}{M(\al_s)}\right)^{\frac{1}{2}}
\big\{\,{}^{\al_1}_{\al_3}\,{}^{\al_2}_{\al_4}\,{}^{\al_s}_{\al_t}\big\}_b^{}
\,.
\end{equation}
Indeed, formula \rf{eq2} is a straightforward consequence of
equations \rf{GvsF}, \rf{FvsRW} and \rf{norm} above.

\subsection{6j symbols of ${\rm Diff}(S^1)$}

It is known that Liouville theory is deeply related to the 
representation theory of the group ${\rm Diff}(S^1)$ of diffeomorphisms
of the unit circle \cite{T08}. The operator product expansion from conformal 
field theory leads to the definition of a suitable generalization of 
the tensor product operation for representations of infinite-dimensional
groups like ${\rm Diff}(S^1)$. One may therefore 
interpret the chiral vertex-operators
from conformal field theory as analogs of the Clebsch-Gordan
maps, and the fusion coefficients as analog of $6j$-symbols 
\cite{MS,Teschner:2001rv,T08}. 

A similar issue arises here as pointed out above in our discussion 
of the modular double: To find particularly natural normalization
conditions. The normalization defined in \rf{Blnorm} above, while
being
natural from the physical point of view,
is not a direct counterpart of the normalization condition
used to define the $6j$ symbols of the modular double above.
Such a normalization condition can naturally be defined by 
requiring invariance under the Weyl-reflections $\al_i\ra Q-\al_i$.
Due to the factors $\Upsilon(2\al_i)$ 
in the definition of $C(\al_3,\al_2,\al_1)$, 
the  symmetry under  $\al_i\ra Q-\al_i$
is spoiled by the change of normalization  
\rf{Blnorm}.

However, it is easy to restore this symmetry by replacing the
normalization factor $C(\al_3,\al_2,\al_1)$ entering the
definition \rf{Blnorm} by 
\begin{align}
D(\al_1 &,\al_2,\al_3)  = \frac{|\Ga_b(2\al_1)\Ga_b(2\al_2)\Ga_b(2\al_3)|^{-2}}
{\Upsilon(2\al_1)\Upsilon(2\al_2)\Upsilon(2\al_3)}C(\al_3,\al_2,\al_1)\,.
\notag\end{align}
Replacing $C$ by $D$ in \rf{Blnorm} leads to the definition
of normalized conformal blocks $\CK_{\al_s}^{(s)}(A|Z)$
and $\CK_{\al_t}^{(t)}(A|Z)$ which can be interpreted as analogs of 
invariants in tensor products of four representations of
${\rm Diff}(S^1)$. The kernel appearing in the relation  
\begin{equation}
\CK_{\al_s}^{(s)}(A|Z)\,=\,\int_{{Q/2}+\textup{i}\BR} d\mu(\al_t)\;
\big\{\,{}^{\al_1}_{\al_3}\,{}^{\al_2}_{\al_4}\,
{}^{\al_s}_{\al_t}\big\}_{\rm Diff(S^1)}^{}
\CK_{\al_t}^{(t)}(A|Z)\,.
\end{equation}
is naturally interpreted as an analog of
the $6j$ symbols for ${\rm Diff}(S^1)$. It coincides exactly with the
b-$6j$ symbols, 
\begin{equation} \label{eq3}
\big\{\,{}^{\al_1}_{\al_3}\,{}^{\al_2}_{\al_4}\,
{}^{\al_s}_{\al_t}\big\}_{\rm Diff(S^1)}^{}
\,=\,
\big\{\,{}^{\al_1}_{\al_3}\,{}^{\al_2}_{\al_4}\,{}^{\al_s}_{\al_t}\big\}_b^{}
\,.
\end{equation}
as can easily be checked by straightforward calculations.

\section{Application to two-dimensional quantum hyperbolic geometry}
\label{sec:2hyp}

\setcounter{equation}{0}

It is known that the Racah-Wigner symbols of the modular
double play an important role when the 
quantum Teichm\"uller theory \cite{Fo97,Ka97,CF99} 
is studied in the length 
representation \cite{Teschner:2003at,T05}. 
Having fixed a particular
normalization in our definition of the b-$6j$ symbols above
naturally leads to question what it corresponds to in this
context. We are going to show that 
the definition of the b-$6j$ symbols corresponds to 
the quantization of a particular
choice of Darboux-coordinates for the classical Teichm\"uller
spaces.
The Teichm\"uller spaces $\CT(C)$ 
are well-known to be related to a connected
component in the moduli space of flat $SL(2,\BR)$-connections
on Riemann surfaces. Natural
Darboux coordinates for this space have recently been
discussed in \cite{NRS}.

The quantization of the Teichm\"uller spaces will be discussed 
in terms of the Darboux coordinates of \cite{NRS} in a 
self-contained manner in \cite{TV}. In the following we will 
collect some relevant observations that can fairly easily 
be extracted from the existing literature.

\subsection{Classical Teichm\"uller theory of the four-holed sphere}

To be
specific, let us restrict attention to four-holed spheres $C_{0,4}$.
The holes are assumed to be represented by geodesics
with lengths $L=(l_1, \dots,l_4)$.
There are three simple closed curves $\ga_s$, $\ga_t$, and $\ga_u$
encircling pairs of points $(z_1,z_2)$, $(z_2,z_3)$ and $(z_1,z_3)$,
respectively. A set of useful coordinate functions are defined
in terms of the hyperbolic cosines
$L_\si=2\cosh\frac{l_\si}{2}$, $\si\in\{s,t,u\}$, 
of the geodesic length functions
$l_\si$ on $\CT_{0,4}\equiv \CT(C_{0,4})$. $l_\si$ is defined
as the length of the geodesic $\ga_\si$, defined
by means of the constant negative
curvature metric on $C_{0,4}$. 

The well-known relations
between Teichm\"uller spaces $\CT(C)$ 
and the moduli spaces $\CM_G(C)$ of 
flat $G=SL(2,\BR)$-connections on Riemann surfaces imply 
that the geodesic length functions $L_\si$ are related to the 
holonomies $g_\si$ along $\ga_\si$ as $L_{\si}=-{\rm Tr}(g_\si)$.
This allows us to use the description given in \cite{NRS}, 
which may be briefly summarized as follows. The structure of
$\CM_G(C_{0,4})$ as an algebraic variety is expressed by the fact
that the three
coordinate functions $L_s$, $L_t$ and $L_u$ satisfy one algebraic
relation of the form $\CP_L(L_s,L_t,L_t)=0$. 
The Poisson bracket $\{L_{\si_1},L_{\si_2}\}$ defined
by the Weil-Petersson symplectic form is also algebraic in 
the length variables $L_\si$, and
can be written elegantly in the form 
\begin{equation}\label{loopPB}
\{\,L_s\,,L_t\,\}\,=\,\frac{\pa}{\pa L_u}\CP_L(L_s,L_t,L_t)\,.
\end{equation}
As shown in \cite{NRS} one may represent $L_s$, $L_t$ and $L_u$ 
in terms of Darboux-coordinates $l_s$ and $k_s$ which have
Poisson bracket $\{l_s,k_s\}=2$. The expressions for $L_s$ and $L_t$
are, in particular,
\begin{align}\label{FN-FK}
& L_s\,=\,2\cosh(l_s/2)\,,\\
& L_t(L_s^2-4)\,=\,
2({L}_2L_3+L_1L_4)+L_s(L_1L_3+L_2L_4) 
+2\cosh(k_s)
\sqrt{c_{12}(L_s)c_{34}(L_s)}\,,
\notag\end{align}
where $L_i=2\cosh\frac{l_i}{2}$, and $c_{ij}(L_s)$ is defined as
\begin{align}\label{cijdef}
c_{ij}(L_s) & \,=\,L_s^2+L_i^2+L_j^2+L_sL_iL_j-4\ \\ \nonumber & = 
2 \cosh \fr{l_s+l_i+l_j}{4} 2 \cosh \fr{l_s+l_i-l_j}{4} 
2 \cosh \fr{l_s-l_i+l_j}{4} 2 \cosh \fr{l_s-l_i-l_j}{4}.
\end{align}
Together with a similar formula for $L_u$, these expressions
ensure that both the algebraic relation $\CP_L(L_s,L_t,L_t)=0$ and 
the Poisson structure \rf{loopPB} are satisfied.
These Darboux coordinates
are identical to the Fenchel-Nielsen length-twist coordinates
well-known in hyperbolic geometry.\footnote{This can be
inferred from \cite{ALPS}. We thank T. Dimofte for 
pointing this reference out to us}

Similar Darboux coordinates $(l_t,k_t)$ and  $(l_u,k_u)$
can be associated to the 
curves $\ga_t$ and $\ga_u$, respectively.  
The change of coordinates between the Darboux-coordinates 
$(l_s,k_s)$ and  $(l_t,k_t)$
is represented by a generating function $\CS^{st}_L(l_s,l_t)$ 
such that 
\begin{equation}\label{cantrsf}
\frac{\pa}{\pa l_s}\CS^{st}_L(l_s,l_t)\,=\,-k_s\,,\qquad 
\frac{\pa}{\pa l_t}\CS^{st}_L(l_s,l_t)\,=\,k_t\,. 
\end{equation}

Other natural sets of 
Darboux-coordinates $(l_\si,k_\si')$ can be obtained by means of 
canonical transformations $k_\si'=k_\si+f(l_\si)$. By a suitable 
choice of $f(\si)$, one gets Darboux coordinates $(l_s,k_s')$
in which 
the expression for $L_t$ in \rf{FN-FK} is replaced by 
\begin{align}\label{FN-FK'}
L_t(L_s^2-4)\,=\,&
\;2({L}_2L_3+L_1L_4)+L_s(L_1L_3+L_2L_4) \\
& +
2 \cosh \fr{l_s+l_1-l_2}{4} 2 \cosh \fr{l_s+l_2-l_1}{4} 
2 \cosh \fr{l_s+l_3-l_4}{4} 2 \cosh \fr{l_s+l_4-l_3}{4} \;e^{+k_s'}\notag  \\
& +
2 \cosh \fr{l_s+l_1+l_2}{4} 2 \cosh \fr{l_s-l_1-l_2}{4} 
2 \cosh \fr{l_s+l_3+l_4}{4} 2 \cosh \fr{l_s-l_3-l_4}{4} \;e^{-k_s'}
\,. \notag
\end{align}
The Darboux coordinates $(l_s,k_s')$ are equally good to 
represent the Poisson structure of $\CM_G(C_{0,4})$, but they 
have the advantage that the expressions for $L_\si$ do not 
contain square-roots. This will later turn out to be important. 

\subsection{The quantization problem}

The quantum Teichm\"uller theory \cite{Fo97,Ka97,CF99,CF00}
constructs a non-commutative
algebra $\CA_b$ 
deforming the Poisson-algebra of geodesic length functions
on Teichm\"uller space. In the so-called 
length representation \cite{Teschner:2003at,T05} one may construct 
natural representations of this algebra associated to pants 
decompositions of the Riemann surface under consideration. 

For the case under consideration, the aim is to construct a 
one-parameter family
of non-commutative deformations $\CA_b$ 
of the Poisson-algebra
of functions on $\CT_{0,4}\equiv\CT(C_{0,4})$ which has generators
$\CL_s$, $\CL_t$, $\CL_u$ corresponding to the 
functions 
$L_\si$, $\si\in\{s,t,u\}$, respectively. There is one algebraic
relation that should be satisfied
among the three generators $\CL_s$, $\CL_t$, $\CL_u$.

Natural representations $\pi_\si$, $\si\in\{s,t,u\}$, 
of $\CA_b$ by operators on suitable spaces of functions 
$\psi_\si(l_\si)$ can be constructed in terms of the quantum counterparts of
the Darboux variables $l_\si$, $k_\si$, now represented by the
operators
$\sll_\si$, $\sk_\si$ defined as
\begin{equation}
\sll_\si\,\psi_\si(l_\si):=\,l_\si\,
\psi_\si(l_\si)\,,\qquad 
\sk_s\,\psi_\si(l_\si):=4\pi b^2
\frac{1}{\textup{i}}\frac{\pa}{\pa l_s}\psi_\si(l_\si)\,.
\end{equation}
The operator $\pi_\si(\CL_\si)$ acts as 
operator of multiplication in the representation $\pi_\si$,
$\pi_\si(\CL_\si)\equiv 2\cosh(\sll_\si)/2$.
The remaining two generators of $\CA_b$ are then represented 
as difference operators. Considering the representation
$\pi_s$, for example, we will find that $\pi_s(\CL_t)$
can be represented in the form
\begin{equation}\label{Diffop}
\pi_s(\CL_t)\psi_s(l_s)\,=\,\big[ 
D_+(l_s)e^{+\sk_s}+D_0(l_s)+D_-(l_s)e^{-\sk_s}\big]\psi_s(l_s)\,.
\end{equation}
This formula should of course reproduce \rf{FN-FK} or \rf{FN-FK'}  
in the classical 
limit, but due to ordering issues and other
possible quantum corrections it is a priori far from obvious how 
to define the coefficients $D_\ep(l_s)$, $\ep=-,0,+$. 

Note, in particular, that the 
requirement that $\pi_s(\CL_s)$ acts as multiplication 
operator leaves a large freedom. 
A gauge transformation 
\begin{equation}\label{gauge}
\psi_s(l_s)\,=\,e^{\textup{i}\chi(l_s)}\psi_s'(l_s)\,,
\end{equation}
would lead to a representation $\pi_s'$ of the form \rf{Diffop}
with $\sk_s$ replaced by
\begin{equation}\label{q-can}
\sk'_s:=\,\sk_s+4\pi b^2\,\pa_{l_s}\chi(l_s)\,.
\end{equation}
This is nothing but the quantum version of a
canonical transformation $(l_s,k_s)\ra (l_s,k_s+f(l_s))$. 
The representation 
$\pi_s'(\CL_t)$ may then be obtained from \rf{Diffop}
by replacing 
$D_\ep(l_s)\ra E_\ep(l_s)$ with 
$E_\ep(l_s)$ equal to $e^{\textup{i}(\chi(l_s-4\ep \textup{i} \pi b^2)-\chi(l_s))}
D_\ep(l_s)$ for $\ep=-1,0,1$. Fixing a particular set of Darboux coordinates
corresponds to fixing a particular choice of the
coefficients $D_\ep(l_s)$ in \rf{Diffop}.

\subsection{Transitions between representation}

The transition between any pair of representations 
$\pi_{\si_1}$  and $\pi_{\si_2}$
can be represented as
an integral transformation of the form
\begin{equation}\label{duality}
\psi_{\si_1}(l_{\si_1})\,=\,\int dl_{\si_2}\;A^{\si_1\si_2}_L(l_{\si_1},
l_{\si_2})
\,\psi_{\si_2}(l_{\si_2})\,.
\end{equation}
The relations
\begin{equation}\label{duality'}
\begin{aligned}
& \big(\pi_s(\sk_t)\psi_{s}\big)(l_{s})\,=\,4\pi b^2
  \int dl_{t}\;A^{st}_L(l_{s},l_{t})
\,\frac{1}{\textup{i}}\frac{\pa}{\pa l_t}\psi_{t}(l_{t})\,,\\
& 4\pi b^2\frac{1}{\textup{i}}\frac{\pa}{\pa l_s}
\psi_{s}(l_{s})\,=\,\int dl_{t}\;A^{st}_L(l_{s},
l_{t})
\,\big(\pi_t(\sk_s)\psi_{t}\big)(l_{t})\,,
\end{aligned}
\end{equation}
describing
the quantum change of Darboux coordinates
are direct consequences.

It is important to note that the problem to  find the 
proper quantum representation of the generators
$\pi_\si(\CL_{\si'})$ is essentially equivalent to the problem
to find the kernels $A^{\si_1\si_2}_L(l_{\si_1},
l_{\si_2})$ in \rf{duality}.
Indeed, the requirement that
$\pi_\si(\CL_\si)\equiv 2\cosh(\sll_\si)/2$ implies
difference equations for the kernel $A^{\si_1\si_2}_L(l_{\si_1},
l_{\si_2})$ such as
\begin{equation}\label{Loopduality}
\pi_{\si_1}(\CL_{\si_2})\cdot A^{\si_1\si_2}_L(l_{\si_1},
l_{\si_2})\,=\,2\cosh(l_{\si_2}/2)\,
A^{\si_1\si_2}_L(l_{\si_1},l_{\si_2})\,.
\end{equation}
The difference operator on the left is of course understood
to act on the variable $l_{\si_1}$ only. 
Under certain natural conditions one may
show that the difference equations \rf{Loopduality} 
determine the
kernels $A^{\si_1\si_2}_L(l_{\si_1},l_{\si_2})$
uniquely. Conversely, knowing $A^{\si_1\si_2}_L(l_{\si_1},l_{\si_2})$,
one may show \cite{TV} that it satisfies relations of the 
form \rf{Loopduality}, and thereby deduce the explicit form of 
$\pi_{\si_1}(\CL_{\si_2})$. 

Considering the generalization to Riemann spheres $C_{0,n}$ 
with more than 
four holes it is natural to demand that the full theory can be 
built in a uniform manner
from the local pieces associated to the four-holed spheres 
that appear in a pants decomposition of $C_{0,n}$. This leads to 
severe restrictions on the kernels $A^{st}_L(l_{s},l_{t})$ 
known as the pentagon- and hexagon equations \cite{T05}. We claim 
that the resulting constraints determine $A^{st}_L(l_{s},l_{t})$ 
essentially uniquely up to changes of the normalization associated to 
pairs of pants.

Solutions of these conditions are clearly given by 
the b-$6j$-symbols. It is important to note, however, that a
change of normalization of the form \rf{norm} will be equivalent to
a gauge transformation \rf{gauge}. This means that 
different normalizations of the b-$6j$ symbols are 
in one-to-one correspondence with choices of Darboux-coordinates 
$(l_\si',k_\si')$
obtained from $(l_\si,k_\si)$ by  
canonical transformations of the form $l_\si'=l_\si$, 
$k_\si'=k_\si+f(l_\si)$. Only a very particular normalization
for the b-$6j$ symbols can correspond to the quantization of 
the Fenchel-Nielsen coordinates.

\subsection{Quantization of Fenchel-Nielsen coordinates}

The main observation we want to make here may be summarized in the
following two statements:

\noindent
{\it 1) The geodesic
length operators can be represented in terms of the quantized
Fenchel-Nielsen coordinates as follows:
\begin{subequations}\begin{align}
\pi_s^{\rm can}(\CL_s)\,=\,& 2\cosh(\sll_s/2)\,,\\
\pi_s^{\rm can}(\CL_t)\,=\,&\frac{1}{ 2(\cosh \sll_s - \cos 2\pi b^2)}
\Big(2\cos\pi b^2(L_2L_3+L_1L_4)+
\SL_s (L_1L_3+L_2L_4)\Big)\nonumber\\ \label{quantum't Hooft}
& \quad +  \frac{1}{\sqrt{2\sinh(\sll_{s}/2)}}
{e^{+\sk_s/2}}
\frac{\sqrt{c_{12}(\SL_s)c_{34}(\SL_s)}}{2\sinh(\sll_s/2)}
{e^{+\sk_s/2}}
\frac{1}{\sqrt{2\sinh(\sll_s/2)}}\\[-.5ex]
& \quad +  \frac{1}{\sqrt{2\sinh(\sll_s/2)}}
{e^{-\sk_s/2}}
\frac{\sqrt{c_{12}(\SL_s)c_{34}(\SL_s)}}{2\sinh(\sll_s/2)}
{e^{-\sk_s/2}}
\frac{1}{\sqrt{2\sinh(\sll_s/2)}}\,,
\notag
\end{align}
\end{subequations}
where $\SL_s=2\cosh(\sll_s/2)\equiv\pi_s(\CL_s)$ and $c_{ij}(L_s)$
was defined in \rf{cijdef}. 
The formulae defining the other representations $\pi_t$ and $\pi_u$
are obtained by simple permutations of indices.

\noindent 2)
The kernel describing the transition between representation $\pi_s$ and 
$\pi_t$ is given in terms of the b-$6j$ symbols as 
\begin{equation} \label{Avs6j}
A_L^{st}(l_s,l_t)
\,=\,
\left({{M(\al_t)}{M(\al_s)}}\right)^{\frac{1}{2}}
\big\{\,{}^{\al_1}_{\al_3}\,{}^{\al_2}_{\al_4}\,{}^{\al_s}_{\al_t}\big\}_b^{}
\,\quad{\rm if}\quad \al_i=\frac{Q}{2}+\textup{i}\frac{l_i}{4\pi b}\,,
\end{equation}
for $i=1,2,3,4,s,t$. The formulae for other pairs of representations
are again found by permutations of indices.
}

The relations between Liouville theory and quantum Teichm\"uller theory
found in \cite{Teschner:2003at} allow one to 
shortcut the forthcoming self-contained derivation \cite{TV}
of the claims above.
In \cite{Teschner:2003at} it was found in particular that the conformal blocks 
$\CF_{\al_s}^{(s)}(A|Z)$ represent particular wave-functions 
in some
representation $\pi_s^{\rm Liou}$,
\begin{equation}
\psi_s(l_s)\,=\,\CF_{\al_s}^{(s)}(A|Z)\quad{\rm if}\quad
\al_s=\frac{Q}{2}+\textup{i}\frac{l_s}{4\pi b}\,.
\end{equation}
This relation fixes a specific representation $\pi_s^{\rm Liou}$.
The generator
$\CL_t$ is represented in $\pi_s^{\rm Liou}$ as in \rf{Diffop} with
coefficients $D_\ep^{\rm Liou}(l_s)$ that can be extracted from 
\cite{AGGTV,DGOT}\footnote{Our generator
$\CL_t$ corresponds to $2\cos(\pi bQ)\CL(\ga_{2,0})$ in \cite{DGOT}.}.
Redefining the conformal blocks as in \rf{Blnorm}
is equivalent
to a gauge transformation \rf{gauge}
which transforms the representation 
$\pi_s^{\rm Liou}$ to the 
representation denoted $\pi_s^{\rm can}$. 
It is straightforward to  
calculate the coefficients $D_{\ep}(l_s)$ from 
$D_\ep^{\rm Liou}(l_s)$ using \rf{Blnorm} and 
\rf{DOZZ}. A related observation was recently made in \cite{IOT}. 
The case of the one-holed torus was discussed along similar lines in 
\cite{DiGu}.

Other normalizations for the b-$6j$ symbols will 
correspond to different choices of Darboux-coordinates. In 
the normalization used in \cite{DGOT}, for example, one would find
\begin{subequations}
\begin{align}
\pi_s'(\CL_t)\,=\,&\frac{1}{ 2(\cosh \sll_s - \cos 2\pi b^2)}
\Big(2\cos\pi b^2(L_2L_3+L_1L_4)+
\SL_s (L_1L_3+L_2L_4)\Big)\nonumber\\ \label{quantum't HooftDGOT}
& +  \frac{4}{\sinh(\sll_{s}/2)}
{e^{+\sk_s'/2}}
\frac{\cosh \fr{\sll_s+l_1-l_2}{4} \cosh \fr{\sll_s+l_2-l_1}{4} 
\cosh \fr{\sll_s+l_3-l_4}{4} \cosh \fr{\sll_s+l_4-l_3}{4}}{\sinh(\sll_s/2)}
{e^{+\sk_s'/2}}
\notag \\
& +  \frac{4}{{\sinh(\sll_s/2)}}
{e^{-\sk_s'/2}}
\frac{\cosh \fr{\sll_s+l_1+l_2}{4} \cosh \fr{\sll_s-l_1-l_2}{4} 
\cosh \fr{\sll_s+l_3+l_4}{4} \cosh \fr{\sll_s-l_3-l_4}{4}}{\sinh(\sll_s/2)}
{e^{-\sk_s'/2}}
\,.
\notag
\end{align}
\end{subequations}
As the analytic properties of the coefficients $D_\ep(l_s)$ in \rf{Diffop}
are linked with the analytic properties of the kernels $A^{st}_L(l_{s},l_{t})$
via \rf{Loopduality},
it is no surprise that the kernels $A'^{st}_L(l_{s},l_{t})$ 
associated to the representation $\pi_s'$
have much better analytic properties than 
$A^{st}_L(l_{s},l_{t})$ as given by 
\rf{Avs6j}.
One may see see these analytic properties as
a profound consequence of the structure of the
moduli spaces $\CM_G(C)$ as algebraic varieties.  

\subsection{Classical limit}

The classical counterpart of the expression \rf{quantum't Hooft}
is found by replacing $\sll_s$ and $\sk_s$
by commuting variables $l_s$ and $k_s$, respectively, and 
sending $b\ra 0$. The formulae
for the operators $\pi_s^{\rm can}(\SL_s)$ 
and $\pi_s^{\rm can}(\SL_t)$ given above are thereby 
found to be related to the 
formulae \rf{FN-FK} for $L_s$ and $L_t$ in terms of the 
Darboux coordinates $l_s$ and $k_s$ for $\CT_{0,4}$.
We conclude that the representation $\pi_s^{\rm can}$ is 
the representation associated to the Darboux 
coordinates discussed in \cite{NRS}. The representation $\pi_s'$
reproduces \rf{FN-FK'}.
 
Furthermore, by analyzing the classical limit of 
the relations the relations \rf{duality'} 
with the help of the saddle-point method one may see 
that
the function $S_L^{st}(l_s,l_t)$ which describes
the leading semiclassical asymptotics of the  
kernel $A_L^{st}(l_s,l_t)$ via
\begin{equation}
A_L^{st}(l_s,l_t)\,=\,\exp\bigg(\frac{1}{4\pi \textup{i} b^2}
S_L^{st}(l_s,l_t)\bigg)
\big(1+\CO(b^2)\big)\,,
\end{equation}
must coincide with the generating function for the canonical transformation
between the Darboux-coordinates 
$(l_s,k_s)$ and  $(l_t,k_t)$. As this function is known \cite{NRS} 
to be equal to the volume of the hyperbolic tetrahedron specified
by the lengths $(l_1,l_2,l_3,l_4,l_s,l_t)$, 
we have found 
a second proof of the statement that the semiclassical 
limit of the b-$6j$ symbols is given by the volume of 
such tetrahedra.

\section{Applications to supersymmetric gauge theories}

\setcounter{equation}{0}

\subsection{Three-dimensional gauge theories on duality walls}

Recently remarkable relations between a certain class $\CS$ of 
$\mathcal{N}=2$ supersymmetric four-dimensional gauge theories 
and two-dimensional conformal field theories
have been discovered in \cite{Alday:2009aq}. 
One of the simplest examples for such relations are relations
between the partition functions of certain gauge theories on $S^4$ 
\cite{Pe}
and physical correlation functions in Liouville theory. 
The partition function of the
$\mathcal{N}=2$ SYM theory with $SU(2)$ gauge group and $N_f=4$ 
hypermultiplets, for example, has a very simple expression in terms
of the four-point function \rf{s-channel} in Liouville theory.
The partition function of the $S$-dual theory would be given by
the four-point function \rf{t-channel}, and the equality between
the two expressions \cite{Teschner:2001rv} 
represents a highly nontrivial check of 
the $S$-duality conjecture.

Interesting generalizations of such relations were 
recently suggested in \cite{Drukker:2010jp}: one may consider
two four-dimensional theories from class $\CS$ on the upper- and
lower semispheres of $S^4$, respectively, 
coupled to a three-dimensional theory
on the defect $S^3$ separating the two semi-spheres. 
Choosing the two theories to be the $N_f=4$ theory and
its $S$-dual, for example, the arguments from \cite{Drukker:2010jp}
suggest that the partition function of the full theory
should be given by an expression of the form 
\begin{equation}\label{dualwall}
\int_{({Q}/{2}+\textup{i}\BR)^2}d\al_sd\al_t\;
(\CG_{\al_s}^{(s)}(A|Z))^*\,G_{\al_s\al_t}
\big[\,{}^{\al_3}_{\al_4}\,{}^{\al_2}_{\al_1}\,\big]\,
\CG_{\al_t}^{(t)}(A|{Z}')\,,
\end{equation}
using the notations from Section \ref{Liou}.
The interpretation in terms of 
two four-dimensional theories coupled by a defect
suggests  \cite{Drukker:2010jp} that the kernel $G_{\al_s\al_t}
\big[\,{}^{\al_3}_{\al_4}\,{}^{\al_2}_{\al_1}\,\big]$
in \rf{dualwall} can be interpreted as the partition 
function of a three-dimensional supersymmetric gauge theory
on $S^3$ which represents a boundary condition for 
both of the four-dimensional gauge theories on the 
semi-spheres of $S^4$. 

The identification of the three-dimensional gauge theories 
living on the duality walls may be seen as 
part of a larger program \cite{Terashima:2011qi,DiGu,Dimofte:2011ju}
which aims to develop a three-dimensional
version of the relations discovered in \cite{Alday:2009aq}. 
Roughly speaking, 
the idea is that there should exist a duality between certain
families of three-dimensional supersymmetric gauge theories
and Chern-Simons theories on suitable three-manifolds. 
A procedure was described in \cite{Dimofte:2011ju}
for the geometric construction of 
relevant three-dimensional gauge theories from simple 
building  blocks associated to ideal tetrahedra. 

In the simpler case where the $N_f=4$ theory is replaced by the
$\CN=4$-supersymmetric gauge theory, an ansatz for the relevant 
three-dimensional theory was suggested by the work 
\cite{Gaiotto:2008ak}, where this theory was called $T[SU(2)]$.
In subsequent work \cite{Hosomichi:2010vh,Hama:2010av} it was 
explicit checked that the analog of the kernel $G_{\al_s\al_t}$
for this case is given by the partition 
function of the $T[SU(2)]$ theory. A natural mass-deformation exists for
the $T[SU(2)]$-theory, and it was also shown in 
\cite{Hosomichi:2010vh,Hama:2010av} that its partition 
function would essentially coincide with the counterpart of the 
kernel which would appear in the case of the 
so-called $\mathcal{N}=2^*$-theory rather than 
the $N_f=4$-theory.
However, so 
far no three-dimensional gauge theory which would have the 
b-$6j$ symbols as its partition function 
has been identified yet.

\subsection{Partition functions of three-dimensional supersymmetric 
gauge theories}

Let us briefly review the general form of the 
partition functions for $3d$ supersymmetric field theories.
According to \cite{Hama:2010av}, following \cite{Kapustin:2010xq,J,HHL1},
the partition function for $3d$ $\mathcal{N}=2$ SYM  theory with gauge group $G$ and flavor symmetry group $F$ defined on a squashed three sphere has the form
\beq \label{PF_def}
Z(\underline{f}) \ = \ \int_{-\textup{i}\infty}^{\textup{i}\infty}
\prod_{j=1}^{\rank G}du_j\,  J(\underline{u})
Z^{vec}(\underline{u}) \prod_{I} Z_{\Phi_I}^{chir}(\underline{f},\underline{u}).
\eeq
Here $f_k$ are the chemical potentials for the flavor symmetry group $F$ while
$u_j$-variables are associated with the Weyl weights for the Cartan
subalgebra of the gauge group $G$. For Chern-Simons theories one has
$J(\underline{u})=e^{-\pi \textup{i} k \sum_{j=1}^{\rank G} u_j^2}$,
where $k$ is the level of CS-term, and for SYM theories one has
$J(\underline{u}) = e^{2 \pi \textup{i} \lambda \sum_{j=1}^{\rank G} u_j}$,
where $\lambda$ is the Fayet-Illiopoulos term. There are two different contributions to the partition function (\ref{PF_def}): $Z^{vec}(\underline{u})$ which comes from vector superfields and $Z_{\Phi_I}^{chir}(\underline{f},\underline{u})$ arising from the matter fields. All these terms are expressed in terms of noncompact quantum dilogarithms. The contribution of vector superfield for $G=SU(2)$ which we are interested in coincides
with the Plancherel measure (\ref{Pmeas}) introduced above, 
\begin{equation}
Z^{vec}(\underline{u}) \,=\,M(Q/2 + \textup{i} u)\,,
\end{equation}
as follows from \cite[Equation (5.33)]{Hama:2010av} 
using (\ref{Sb2}) and (\ref{Sb3}). For each chiral superfield $\Phi_I$ 
the contribution to the partition function is $S_b(\alpha)$ where 
$\alpha$ is some linear combination of the $R$-charge and mass 
parameters which can be derived from the group representation of 
the matter content (see, for example, \cite{DSV}).

\subsection{The b-6j symbols as a partition function}

Although expression (\ref{6j2}) for b-$6j$ symbol 
resembles the partition functions of 
$3d$ SYM theory with $U(1)$ gauge group, it cannot easily be interpreted 
as partition function for some three-dimensional 
gauge theory since the parameters 
entering its expression are subject to the condition that 
their sum equals $2Q$, 
while the parameters entering partitions functions are not restricted. 

In the course of the derivation of the new formula 
\rf{6j2} for the b-$6j$ symbols, as described in 
Appendix \ref{Cor}, we have found a few other 
integral representations
for these objects, including
\begin{align} \label{PF-6quarks}
 \mathcal{A}_1 \,I \bigg( 
\begin{array}{ccc}
\fr{Q-\alpha_t-\alpha_1-\alpha_4}{2} + \alpha_s & \fr{3Q-\alpha_t-\alpha_1-\alpha_4}{2}-\alpha_s & \fr{Q+\alpha_1-\alpha_4+\alpha_t}{2}-\alpha_3 \\  \fr{-Q-\alpha_1+\alpha_4+\alpha_t}{2}+\alpha_2 & \fr{Q-\alpha_1+\alpha_4+\alpha_t}{2} - \alpha_2& \fr{-Q+\alpha_1-\alpha_4+\alpha_t}{2}+\alpha_3
\end{array}\bigg)\,,
\end{align}
where we define the integral $I(\underline{\mu})$ as
\begin{equation}
I(\underline{\mu}) = \frac 12 \int_{- \textup{i} \infty}^{\textup{i} \infty} \frac{\prod_{i=1}^6 S_b(\mu_i\pm u)}{S_b(\pm 2u)} du\,,\qquad
[ \underline{\mu} ] = \bigg[  \begin{array}{ccc}
\mu_1 & \mu_2 & \mu_3 \\ \mu_4 & \mu_5 & \mu_6 \end{array}
\bigg]\,,
\end{equation}
and the prefactor in \rf{PF-6quarks} is explicitly given as
\beq \nonumber
\mathcal{A}_1 = \frac{S_b(\alpha_2+\alpha_3-\alpha_t) S_b(\alpha_1-\alpha_2+\alpha_s) S_b(-Q+\alpha_1+\alpha_4+\alpha_t)}{S_b(\alpha_2+\alpha_t-\alpha_3) S_b(\alpha_3+\alpha_t-\alpha_2) S_b(\alpha_3-\alpha_4+\alpha_s)}.
\eeq
We would like to point out that this expression, as opposed to \rf{6j2},
admits an interpretation as a partition 
function of the form (\ref{PF_def}) for a certain three-dimensional 
SYM theory. 
Namely, the expression \rf{PF-6quarks} without coefficient 
$\mathcal{A}_1$ can be interpreted as the partition function 
of three-dimensional $\mathcal{N}=2$ SYM theory defined on a 
squashed three-sphere with $SU(2)$ gauge group and $6$ quarks 
in the fundamental representation of the gauge group. The flavor symmetry 
group is $SU(6) \times U(1)_A \times U(1)_R$. The total axial mass is $m_A=\frac 16 \sum_{i=1}^6 \mu_6$ while the masses of 6 chiral multiplets then is $m_i = \mu_i - \frac 16 \sum_{k=1}^6 \mu_k,i=1,\ldots,6$ 
(constrained to $\sum_{i=1}^6 m_i=0$). We also 
take the $R$-charge in UV to be 0.
Considering \rf{PF-6quarks} 
as the partition function for $3d$ $\mathcal{N}=2$ SYM theory one obtains 
a whole series of Seiberg dualities which can be derived from \cite{DSV} 
by taking $N=1$ there. Keeping in mind the coefficient $\mathcal{A}_1$ 
in \rf{PF-6quarks} one sees that the corresponding theory has $8$ more 
singlet chiral fields and the flavor symmetry group is broken to 
$U(1)^5 \times U(1)_A \times U(1)_R$. 

We would also like to remark that the identification of the b-$6j$ symbols
as partition functions works straightforwardly 
only for the b-$6j$ symbol
$\big\{\,{}^{\al_1}_{\al_3}\,{}^{\al_2}_{{\al}_4}\,|\,{}^{\al_s}_{\al_t}\big\}_b^{\rm an}$. The square-roots
appearing in the expression for 
$\big\{\,{}^{\al_1}_{\al_3}\,{}^{\al_2}_{{\al}_4}\,|\,{}^{\al_s}_{\al_t}\big\}_b$ seem to prevent a similar interpretation.

\subsection{Applications to the geometric construction of three-dimensional
gauge theories?}

It is interesting to observe that the result for (mass-deformed)
$T[SU(2)]$ from \cite{Hosomichi:2010vh,Hama:2010av}, 
after applications of the same type 
of identities, can be brought to $3d$ $\mathcal{N}=2$ CS theory with 
$SU(2)$ gauge group at level $1$, $4$ quarks and some singlet chiral fields. The above statement can be derived from the following integral identity \cite{SV}
\begin{align}
& \int_{-\textup{i} \infty}^{\textup{i} \infty} \frac{
S_b({Q}/{4} - \mu+{m}/{2} \pm z)}{
S_b({3Q}/{4} - \mu-{m}/{2} \pm z)} e^{4 \pi \textup{i} \xi z} dz 
\\ & \;\;\;
= \frac 12 e^{2 \pi \textup{i} (\xi^2-(\frac Q4 +\frac m2)^2+\mu^2)} 
S_b(Q/2 - m \pm 2 \xi)
\int_{-\textup{i} \infty}^{\textup{i} \infty} 
\frac{S_b(\frac{Q}{4}+\frac{m}{2} \pm \mu \pm \xi \pm y)}
{S_b(\pm 2 y)} e^{-2 \pi \textup{i} y^2} dy\,.
\notag\end{align}
These two observations suggest that there may be an analog
of the geometric construction of three-dimensional supersymmetric 
gauge theories discussed in \cite{Dimofte:2011ju} 
which is 
based on building blocks with $SU(2)$ gauge symmetry rather
than $U(1)$ gauge symmetry.
Indeed, the two three-dimensional 
partition functions discussed above
can be identified with the kernels for the
fusion move $A$ and for the modular transformation of the one-punctured
torus $S$, respectively. Together with the braiding, the two 
kernels above generate a representation of the modular groupoid
\cite{T08}. This is what one needs to apply standard 
methods for the combinatorial quantization of Chern-Simons theories
to the case of $SL(2,\BR)$-Chern-Simons theory.
It is also suggestive to point out that the number of 
quarks of the theory whose partition function gives \rf{PF-6quarks}
nicely matches with the number of  
angles defining the generic hyperbolic tetrahedron. 

We take these
observations above as a hint that three-dimensional $\mathcal{N}=2$ SYM 
theory with $SU(2)$ gauge group and $6$ quarks plus some number of 
singlets could be associated to the non-ideal hyperbolic tetrahedron
in a future generalization of the constructions in 
\cite{Dimofte:2011ju}, 
where the triangulations of 
three-manifold by
ideal tetrahedra are replaced by triangulations by 
non-ideal tetrahedra.
This raises several interesting questions  
which should be clarified, including, in particular, 
the interpretation of 
normalization changes for b-$6j$ symbols (\ref{norm}) 
from the point of view of supersymmetric gauge theories.

\noindent
{\bf Acknowledgements}
We would like to thank T. Dimofte, S. Gukov, R. Kashaev and
S. Shatashvili for useful discussions on related topics.

 \newpage

\appendix

\section{Special functions}\label{Qdil}
\setcounter{equation}{0}

\subsection{The function $\Ga_b(x)$}

The function $\Ga_b(x)$ is a close relative of the double
Gamma function studied in \cite{Ba}. It 
can be defined by means of the integral representation
\begin{equation}
\log\Ga_b(x)\;=\;\int\limits_0^{\infty}\frac{dt}{t}
\biggl(\frac{e^{-xt}-e^{-Qt/2}}{(1-e^{-bt})(1-e^{-t/b})}-
\frac{(Q-2x)^2}{8e^t}-\frac{Q-2x}{t}\biggl)\;\;.
\end{equation}
Important properties of $\Ga_b(x)$ are
\begin{align}
{} \text{functional equation} \quad &
\Ga_b(x+b)=\sqrt{2\pi}b^{bx-\frac{1}{2}}\Ga^{-1}(bx)\Ga(x). \label{Ga_feq}\\
{}\text{analyticity}\quad &
\Ga_b(x)\;\text{is meromorphic,}\nonumber\\ 
{}&\hspace{2.5cm}\text{poles:}\;\,  
x=-nb-mb^{-1}, n,m\in\BZ^{\geq 0}. 
\end{align}
A useful reference for further properties is \cite{Sp}.

\subsection{Double Sine function}\label{Qdil1}

The special functions used in this note are all
build from the so-called double Sine-function. This function
is closely related to the special function here denoted 
$e_b(x)$, which was introduced under the name of {\em quantum dilogarithm}
in \cite{FK2}. These special functions are simply related
to the Barnes double Gamma function \cite{Ba}, and were also introduced 
in studies of quantum groups and integrable models in 
\cite{F2,Ru,Wo,V}.

In the strip $|{\rm Im}(x)| < \fr{Q}{2}$, function $e_b(x)$ has the 
following integral representation
\begin{equation}\label{wint}
 e_b(x)= 
 \exp \Biggl\{ -
 \int\limits_{\BR+\textup{i}0} \frac{dt}{4\, t} \,
 \frac{ e^{-2\textup{i} t x}}{\sinh b t \, \sinh{\frac{t}{b}} } \Biggr\} \,,
\end{equation}
where the integration contour goes around the pole $t=0$ in the 
upper half--plane. 
The function $s_b(x)$ is then 
related to $e_b(x)$ as follows
\begin{equation}
 s_b(x)\,=\, 
 e^{\frac{\textup{i} \pi}{2} x^2 + 
      \frac{\textup{i} \pi}{24}(b^2 + b^{-2})}e_b(x)\,.
\end{equation}
The analytic continuation of $s_b(x)$ to the
entire complex plane is a meromorphic function with the
following properties 
\begin{align}
\text{functional equation}  \label{wfunrel} \quad&
 \frac{s_b(x + \fr{\textup{i}}{2}b^{\pm 1})}{s_b(x - \fr{\textup{i}}{2}b^{\pm 1})} = 
 2 \, \cosh (\pi b^{\pm 1} x) \,, \\[0.5mm]
\text{reflection property} \label{wrefl} \quad&
 s_b(x) \; s_b(-x) = 1  \,, \\[0.5mm]
  \label{wcc} \text{complex conjugation} \quad&
  \overline{s_b(x)} = s_b(-\bar{x}) \,,\\[0.5mm]
 \label{wan} 
 \text{zeros\,/\,poles} \quad& 
 (s_b(x))^{\pm 1} = 0 \ \Leftrightarrow  
  \pm x \in  \big\{ \textup{i}\fr{Q}{2}{+}nb{+}mb^{-1};n,m\in\BZ^{\geq 0}\big\}   
	\,, \\[0.5mm]
\label{wres}
 \text{residue} \quad& 
  \Res_{x=-\textup{i}\frac{Q}{2}} s_b(x)=\frac{\textup{i}}{2\pi}
 \,, \\[0.5mm]
 \text{asymptotics} \quad& s_b(x) \sim 
\left\{
\begin{aligned}
& e^{- \frac{\textup{i} \pi}{2}(x^2 + \frac{1}{12}(b^2+b^{-2}))}\;\;{\rm for}\;\,
|x|\ra\infty,\;\, |{\rm  arg}(x)|<\fr{\pi}{2} \,, \\
 & e^{+ \frac{\textup{i} \pi}{2}(x^2 + \frac{1}{12}(b^2+b^{-2}))}\;\;{\rm for}\;\,
|x|\ra\infty,\;\, |{\rm  arg}(x)|>\fr{\pi}{2} \,.
\end{aligned}\right.
\end{align}
Of particular importance for us is the behavior for $b\ra 0$, which 
is given as 
\begin{equation}
e_b\left(\frac{v}{2\pi b}\right)\,=\,
\exp\bigg(\!\!-\frac{1}{2\pi b^2}\Li_2(-e^v)\bigg)\Big(1+\CO(b^2)\Big)\,.
\end{equation}

In our paper we mainly use 
the special function $S_b(x)$ defined by
\begin{equation}\label{Sbx}
 S_b(x) := s_b(\textup{i}x -\fr{\textup{i}}{2}Q) 
\end{equation}
and has the properties
\begin{align}
\label{Sb1}
\text{self--duality} \quad&  
 S_b(x) = S_{b^{-1}}(x) \,, \\
\label{Sb2}
\text{functional equation} \quad&
 S_b(x + b^{\pm 1}) = 2 \, \sin (\pi b^{\pm 1} x) \, S_b(x) \,,\\
\label{Sb3}
\text{reflection property} \quad&
  S_b(x) \, S_b(Q-x) = 1 \,.
\end{align}
The behavior of $S_b(x)$ for $b\ra 0$
is then given as 
\begin{equation}\label{Sbscl}
S_b\left(\frac{\nu}{2\pi b}\right)\,=\,e^{-\frac{\textup{i}}{2\pi b^2}
(\frac{1}{4}\nu^2-\frac{\pi}{2}\nu+\frac{1}{6}\pi^2)}
\exp\bigg(\!\!-\frac{1}{2\pi \textup{i} b^2}
\Li_2(e^{\textup{i}\nu})\bigg)\Big(1+\CO(b^2)\Big)\,.
\end{equation}
In terms of $\Ga_b(x)$ the double Sine-function is given as
$$
S_b(x) = \frac{\Ga_b(x)}{\Ga_b(Q-x)}.
$$

\subsection{The elliptic Gamma function}
The second class of special functions we need here is the elliptic gamma function which appeared implicitly in \cite{Bx} and was introduced in \cite{Ru}
\begin{equation}\label{ellgamma}
\Gamma(z;p,q) \ = \ \prod_{i,j=0}^\infty \frac{1-z^{-1}p^{i+1}q^{j+1}}{1-zp^iq^j},
\end{equation}
satisfying the following properties
\begin{align}
\text{symmetry} \label{ellgdual}  \quad& 
 \Gamma(z;p,q) = \Gamma(z;q,p) \,,  \\[0.5mm]
\text{functional equations}  \label{ellgfunrel} \quad&
 \Gamma(qz;p,q) = \theta(z;p) \Gamma(z;p,q), \\
 & \Gamma(pz;p,q) = \theta(z;q) \Gamma(z;p,q) \,, \\[0.5mm]
\text{reflection property} \label{ellgrefl} \quad&
 \Gamma(z;p,q) \; \Gamma(\frac{pq}{z};p,q) = 1  \,, \\[0.5mm]
 \label{ellgz} 
 \text{zeros} \quad& 
 z \in  \big\{ p^{i+1}q^{j+1};i,j\in\BZ^{\geq 0}\big\}   
	\,, \\[0.5mm] \label{ellp} 
 \text{poles} \quad& 
 z \in  \big\{ p^{-i}q^{-j};i,j\in\BZ^{\geq 0}\big\}   
	\,, \\[0.5mm]
\label{ellgres}
 \text{residue} \quad& 
  \Res_{z=1} \Gamma(z;p,q) = -\frac{1}{(p;p)_\infty (q;q)_\infty}.
\end{align}

Here $\theta(z;p)$ is a theta-function
$\theta(z;p) = (z;p)_\infty (pz^{-1};p)_\infty.$

\section{Proof of identity \rf{6j2}}\label{idproof}

\setcounter{equation}{0}

\subsection{The master integral identity}
Let us start from the $V$-function \cite{S2} which is the example from Spiridonov' theory of elliptic hypergeometric integrals \cite{S1,S2}\footnote{From physical point of view this integral is the so-called superconformal index for four-dimensional SQCD theory with $SU(2)$ gauge group and $N_f=4$ flavors. The integral transformations for $V$-function describe the multiple duality effect for the above theory \cite{SV0}.} defined by
\begin{equation}\label{Vfunc}
V(\underline{s}) = \kappa \int_{\mathbb{T}} \frac{\prod_{i=1}^8 \Gamma(s_i z^{\pm1};p,q)}{\Gamma(z^{\pm2};p,q)} \frac{dz}{2 \pi \textup{i} z},
\end{equation}
where $\prod_{i=1}^8 s_i = (pq)^2$ is the so-called balancing condition and $$\kappa = \frac{(p;p)_\infty(q;q)_\infty}{2}$$ with $(z;q)_\infty=\prod_{i=0}^\infty(1-zq^i)$. The main building block is the elliptic gamma function defined in
\rf{ellgamma} above.

\begin{thm}\cite{S2}
\begin{equation}\label{Vtransf}
V(\underline{s}) = \prod_{1 \leq i < j \leq 4} \Gamma(s_is_j;p,q) \Gamma(s_{i+4}s_{j+4};p,q) V(\underline{t}),
\end{equation}
where
\begin{eqnarray} \nonumber 
t_i = \varepsilon s_i, i=1,2,3,4; \ \ \ t_i = \varepsilon^{-1} s_i, i=5,6,7,8,
\end{eqnarray}
and $$\varepsilon=\sqrt{\frac{pq}{s_1s_2s_3s_4}}=\sqrt{\frac{s_5s_6s_7s_8}{pq}}.$$
\end{thm}

The integral identities used in this paper will be obtained from \rf{Vtransf}
by limiting procedures \cite{ds:unit} which reduce the elliptic gamma functions
to double Sine functions.
First, we reduce $V$-function to the level of hyperbolic 
$q$-hypergeometric integrals
using the reparameterization of variables
\begin{equation}
z = e^{2 \pi \textup{i} r u}, \qquad s_i = e^{2 \pi \textup{i} r \mu_i},
\quad i=1,\ldots,8, \qquad
p = e^{2 \pi \textup{i} b r}, \qquad q = e^{2 \pi \textup{i} r/b},
\label{limit_r}\end{equation}
and the subsequent limit $r \rightarrow 0$.
In this limit the elliptic gamma function has the following asymptotics
\beq \nonumber
\Gamma(e^{2 \pi \textup{i} r z};e^{2 \pi \textup{i} r b},
e^{2 \pi \textup{i} r/b}) \stackreb{=}{r \rightarrow 0}
e^{-\pi \textup{i}(2z-b-1/b)/12r} S_b(z).\eeq
Using it in the reduction, one obtains an integral lying on the top
of a list of integrals emerging as degenerations of the $V$-function
(we omit some simple diverging exponential multiplier appearing in this
limit together with $-\textup{i}$),
\beq \label{top_int}
I_{h}(\mu_1,\ldots,\mu_8) = \frac 12 \int_{- \textup{i} \infty}^{\textup{i} \infty}
\frac{\prod_{i=1}^8 S_b(\mu_i \pm u)}
{S_b(\pm 2u)} du,
\eeq
with the balancing condition $\sum_{i=1}^8 \mu_i = 2 (b+b^{-1})$.
It has the following symmetry transformation formula descending from the elliptic one
\beq \label{top}
I_{h}(\mu_1,\ldots,\mu_8) = \prod_{1 \leq i < j \leq 4}
S_b(\mu_i + \mu_j)
\prod_{5 \leq i < j \leq 8} S_b(\mu_i + \mu_j)
I_{h}(\nu_1,\ldots,\nu_8),
\eeq
where
$
\nu_i = \mu_i + \xi, \,\nu_{i+4} = \mu_{i+4} - \xi,  i=1,2,3,4,
$
and the parameter $\xi$ is
$$2\xi \ = \ \sum_{i=5}^8 \mu_i - b-b^{-1}= b+b^{-1} - \sum_{i=1}^4 \mu_i.$$
Formula \rf{top} will be our main tool in the following.

\subsection{Useful corollaries.} \label{Cor}
For proving the main transformation formula which allows us to get from (\ref{6j1}) the expression (\ref{6j2}) we need following corollaries.

\begin{cor}\label{Corollary 1}
\begin{eqnarray}
I(\underline{\mu}) = S_b(\mu_5+\mu_6) S_b(2Q-\sum_{i=1}^6 \mu_6) \prod_{1 \leq i < j \leq 4} S_b(\mu_i+\mu_j) I(\underline{\nu}),
\end{eqnarray}
where we define the integral $I(\underline{\mu})$ as
\begin{equation}
I(\underline{\mu}) = \frac 12 \int_{- \textup{i} \infty}^{\textup{i} \infty} \frac{\prod_{i=1}^6 S_b(\mu_i\pm u)}{S_b(\pm 2u)} du.
\end{equation}
Here we have
$$
[\nu_1,\nu_2,\nu_3,\nu_4,\nu_5,\nu_6] \ = \ [\mu_1+\xi,\mu_2+\xi,\mu_3+\xi,\mu_4+\xi,\mu_5-\xi,\mu_6-\xi]
$$
and 
$$
2\xi \ = \ Q - \sum_{i=1}^4 \mu_i\,.
$$
\end{cor}
Later it will be convenient to write 6 variables $\underline{\mu}$ in the following way 
\begin{align} \nonumber 
[ \underline{\mu} ] = \bigg[  \begin{array}{ccc}
\mu_1 & \mu_2 & \mu_3 \\ \mu_4 & \mu_5 & \mu_6 \end{array}
\bigg].
\end{align}

\begin{cor}\label{Corollary 2a}:
\begin{eqnarray}
J(\underline{\mu},\underline{\nu}) = \prod_{i=1}^3 S_b(\mu_i+\nu_4) S_b(\nu_i+\mu_4) I(\underline{\rho}),
\end{eqnarray}
with
\begin{equation}
J(\underline{\mu},\underline{\nu}) = \int_{- \textup{i} \infty}^{\textup{i} \infty} \prod_{i=1}^4 S_b(\mu_i - u) S_b(\nu_i + u) du,
\end{equation}
which has $U(1)$ gauge symmetry, and the balancing condition $\sum_{i=1}^4 (\mu_i+\nu_i)=2Q.$
Here we have
$$
[\rho_1,\rho_2,\rho_3,\rho_4,\rho_5,\rho_6] \ = \ [\mu_1+\xi,\mu_2+\xi,\mu_3+\xi,\nu_1-\xi,\nu_2-\xi,\nu_3-\xi]
$$
and 
$$
2\xi \ = \ Q - \nu_4 - \sum_{i=1}^3 \mu_i = -Q + \mu_4 + \sum_{i=1}^3 \nu_i.
$$
\end{cor}
Again it is useful to have the following notation
\begin{align} \nonumber 
[ \underline{\mu}, \underline{\nu} ] = \bigg[  \begin{array}{cccc}
\mu_1 & \mu_2 & \mu_3 & \mu_4 \\ \nu_1 & \nu_2 & \nu_3 & \nu_4 \end{array}
\bigg],
\end{align}

The inversion of Corollary \ref{Corollary 2a} is the following

\begin{cor}\label{Corollary 2b}
\begin{eqnarray}
I(\underline{\rho}) = \prod_{1 \leq i < j \leq 3} S_b(\rho_i+\rho_j) S_b(\rho_{i+3}+\rho_{j+3}) J(\underline{\mu},\underline{\nu}),
\end{eqnarray}
and the balancing condition $\sum_{i=1}^4 \mu_i+\nu_i=2Q.$
Here we have
\begin{align}
& \nonumber \bigg[  \begin{array}{cccc}
\mu_1 & \mu_2 & \mu_3 & \mu_4 \\ \nu_1 & \nu_2 & \nu_3 & \nu_4 \end{array}
\bigg]
= \bigg[\begin{array}{cccc} \rho_1-x & \rho_2-x & \rho_3-x & Q-\rho_{456}-x\\ 
\rho_4+x & \rho_5+x & \rho_6+x & Q-\rho_{123}+x 
\end{array}\bigg],
\end{align}
where $\rho_{123}=\rho_1+\rho_2+\rho_3$, $\rho_{456}=\rho_4+\rho_5+\rho_6$, 
and $x$ is arbitrary.
\end{cor}

\begin{cor}\label{Corollary 3}:
\begin{eqnarray}
I(\underline{\mu}) = S_b(2Q-\sum_{i=1}^6 \mu_i) \prod_{1 \leq i < j \leq 6} S_b(\mu_i+\mu_j) I(Q/2 - \underline{\mu}).
\end{eqnarray}
\end{cor}

To get the desired transformation formulas one should use
the following asymptotic formulas when some of the parameters go to infinity
\beqa
\lim_{u \rightarrow \infty}e^{\frac{\pi \textup{i}}{2} B_{2,2}(u)} S_b(u)
& = & 1, \text{ \ \ for } \text{arg }b < \text{arg } u <
\text{arg }1/b + \pi, \nonumber \\
\lim_{u \rightarrow \infty}e^{-\frac{\pi \textup{i}}{2} B_{2,2}(u)} S_b(u)
& = & 1, \text{  \ \ for } \text{arg }b - \pi < \text{arg } u <
\text{arg }1/b. \nonumber
\eeqa
By taking different restrictions for the parameters one can get lots of identities from the integral identity (\ref{top}). Let us take 
$$\mu_1 \rightarrow \mu_1 + \mu; \ \ \ \mu_5 \rightarrow \mu_5 - \mu$$
with the following limit $\mu \rightarrow \infty$. The left hand-side of (\ref{top}) gives
\beq \label{r1}
I_{h}(\mu_2,\mu_3,\mu_4,\mu_6,\mu_7,\mu_8) = \frac 12 \int_{- \textup{i} \infty}^{\textup{i} \infty}
\frac{\prod_{i=2}^4 S_b(\mu_i \pm z) S_b(\mu_{i+4} \pm z)}
{S_b(\pm 2z)} dz,
\eeq
without any restrictions for parameters $\mu_2,\mu_3,\mu_4,\mu_6,\mu_7,\mu_8$ and in the right hand-side one needs to shift the integration variable $z \rightarrow z-\mu/2$ and afterwards taking the limit $\mu \rightarrow \infty$ which gives
\begin{align} \label{r2}
 \prod_{2 \leq i < j \leq 4}& S_b(\mu_i+\mu_j) S_b(\mu_{i+4}+\mu_{j+4}) \\
\notag &\times \int_{- \textup{i} \infty}^{\textup{i} \infty}\!\!dz\; 
S_b(\xi+z+({\mu_1+\mu_5})/{2} ) 
S_b((\mu_1+\mu_5)/{2} - \xi - z)
\\  & \qquad \qquad \times 
\prod_{i=2}^4 S_b(\mu_i + \xi - (\mu_1+\mu_5)/{2} - z) S_b(\mu_{i+4} - \xi + (\mu_1+\mu_5)/{2} + z) dz,
\notag\end{align}
and $2 \xi = Q - \sum_{i=2}^4 \mu_i$.

Inverting now the equality (\ref{r1})=(\ref{r2}) one gets Corollary \ref{Corollary 2b}. To get Corollary \ref{Corollary 1} one takes the limit $\mu_7,\mu_8 \rightarrow \infty$ such that $\mu_7-\mu_8=O(1)$ in (\ref{top}).

Application of (\ref{Vtransf}) twice and thrice gives new integral transformations formulas for (\ref{Vfunc}) while further application of (\ref{Vtransf}) does not lead to new integral transformations. It can be shown \cite{S3}
\beq
V(s_1,\ldots,s_8) = \prod_{1 \leq i < j \leq 8}
\Gamma(s_is_j;p,q) \,V\left(\frac{\sqrt{pq}}{s_1},\ldots,
\frac{\sqrt{pq}}{s_8}\right)\,,
\eeq
the reduction to the hyperbolic level of which brings to Corollary 
\ref{Corollary 3}.

In \cite{SV} other reductions of $V$-functions were considered in connections with the so-called state integral for $\textbf{4}_1$ knot \cite{Hikami1} and with the kernel of $S$-move \cite{Teschner:2003at}.

\subsection{Derivation of the indentity \rf{6j2}}

Let us start from the expression (\ref{6j1}) and apply 
Corollary \ref{Corollary 2a} taking parameters as
\begin{align} \nonumber 
[ \underline{\mu}, \underline{\nu} ] = \bigg[  \begin{array}{ccc}
Q \pm (\alpha_s - \fr{Q}{2})& \alpha_2 + \alpha_4 + \alpha_t-\fr{Q}{2} & \alpha_2 + \alpha_4 - \alpha_t + \fr{Q}{2}  \\   -\alpha_4 \pm (\alpha_3-\fr{Q}{2})& \fr{Q}{2}-\alpha_1-\alpha_2 & -\fr{Q}{2} +\alpha_1-\alpha_2 \end{array}
\bigg],
\end{align}
one gets 
\begin{align} \label{intermed1}
 \mathcal{A}_1 I \bigg( 
\begin{array}{ccc}
\fr{Q-\alpha_t-\alpha_1-\alpha_4}{2} + \alpha_s & \fr{3Q-\alpha_t-\alpha_1-\alpha_4}{2}-\alpha_s & \fr{Q+\alpha_1-\alpha_4+\alpha_t}{2}-\alpha_3 \\  \fr{-Q-\alpha_1+\alpha_4+\alpha_t}{2}+\alpha_2 & \fr{Q-\alpha_1+\alpha_4+\alpha_t}{2} - \alpha_2& \fr{-Q+\alpha_1-\alpha_4+\alpha_t}{2}+\alpha_3
\end{array}\bigg)
\end{align}
with
\beq \nonumber
\mathcal{A}_1 = \frac{S_b(\alpha_2+\alpha_3-\alpha_t) S_b(\alpha_1-\alpha_2+\alpha_s) S_b(-Q+\alpha_1+\alpha_4+\alpha_t)}{S_b(\pm(Q-2\alpha_t)) S_b(\alpha_2+\alpha_t-\alpha_3) S_b(\alpha_3+\alpha_t-\alpha_2) S_b(\alpha_3-\alpha_4+\alpha_s)}.
\eeq
The integral in \rf{intermed1} is defined for $\al_k\in Q/2+\textup{i}\BR$ by
using a contour $\widetilde{C}$ that approaches $\frac{Q}{4}+\textup{i}\BR$ near infinity,
and passes the real axis in $(-\frac{Q}{4},\frac{Q}{4})$, and for other values 
of $\al_k\in \frac Q2 + \textup{i}\BR$ by analytic continuation.

Applying Corollary  \ref{Corollary 1} to (\ref{intermed1}) (with the order of parameters as staying in (\ref{intermed1})) one obtains
\begin{align} \label{intermed2} 
\mathcal{A}_2 \,I \bigg( 
\begin{array}{ccc}  \alpha_s + \fr{\alpha_3-\alpha_2-\alpha_t}{2} & Q-\alpha_s+\fr{\alpha_3-\alpha_2-\alpha_t}{2}&  \alpha_1+\fr{\alpha_t-\alpha_2-\alpha_3}{2}  \\ \alpha_4 -Q+ \fr{\alpha_2+\alpha_3+\alpha_t}{2}& Q-\alpha_1+\fr{\alpha_t-\alpha_2-\alpha_3}{2}& -\alpha_4+\fr{\alpha_2+\alpha_3+\alpha_t}{2} \end{array}\bigg),\end{align} 
defined by the contour $\widetilde{C}$ and where
\beq \nonumber
\mathcal{A}_2 = \frac{S_b(\alpha_2+\alpha_3-\alpha_t) S_b(-\alpha_1+\alpha_2+\alpha_s) S_b(\alpha_1+\alpha_4-\alpha_t) S_b(2Q-\alpha_3-\alpha_4-\alpha_s)}{S_b(\pm(Q-2\alpha_t)) S_b(\alpha_3-\alpha_4+\alpha_s) S_b(\alpha_3+\alpha_4-\alpha_s)}.
\eeq

On the next step we apply Corollary \ref{Corollary 3} to (\ref{intermed2}) and get
\begin{align} \label{intermed}
\mathcal{A}_3 \,I \left(\begin{array}{ccc} -\alpha_s + \fr{Q+\alpha_2-\alpha_3+\alpha_t}{2}& \alpha_s + \fr{-Q+\alpha_2-\alpha_3+\alpha_t}{2}&-\alpha_1 + \fr{Q+\alpha_2+\alpha_3-\alpha_t}{2}\\ -\alpha_4 + \fr{3Q-\alpha_2-\alpha_3-\alpha_t}{2}& \alpha_1 + \fr{-Q+\alpha_2+\alpha_3-\alpha_t}{2}& \alpha_4 + \fr{Q-\alpha_2-\alpha_3-\alpha_t}{2}\end{array} \right),
\end{align}
with the same contour $\widetilde{C}$ and
\begin{align}
 \mathcal{A}_3 = &\frac{S_b(\alpha_1-\alpha_2+\alpha_s) S_b(\alpha_1-\alpha_4+\alpha_t) S_b(\alpha_1+\alpha_4-\alpha_t)}{S_b(\pm(Q-2\alpha_t)) S_b(\alpha_2-\alpha_3+\alpha_t) S_b(\alpha_1+\alpha_2-\alpha_s) S_b(2Q-\alpha_1-\alpha_4-\alpha_t)} \nonumber \\ &  \times \frac{S_b(2Q-\alpha_1-\alpha_2-\alpha_s) S_b(-\alpha_1+\alpha_4+\alpha_t)}{S_b(-\alpha_3+\alpha_4+\alpha_s) S_b(2Q-\alpha_2-\alpha_3-\alpha_t) S_b(\alpha_1+\alpha_4-\alpha_t) S_b(-\alpha_2+\alpha_3+\alpha_t)}. \nonumber
\end{align}

Finally, we apply Corollary \ref{Corollary 2b} for (\ref{intermed}) with slightly permuted parameters (since the integral has $S_6$ permutation symmetry over parameters)
\begin{align}
\mathcal{A}_3 \,I \bigg(
\begin{array}{ccc}\alpha_s + \fr{-Q+\alpha_2-\alpha_3+\alpha_t}{2} & \alpha_1 + \fr{-Q+\alpha_2+\alpha_3-\alpha_t}{2}& -\alpha_4 + \fr{3Q-\alpha_2-\alpha_3-\alpha_t}{2} \\ -\alpha_s + \fr{Q+\alpha_2-\alpha_3+\alpha_t}{2}& -\alpha_1 + \fr{Q+\alpha_2+\alpha_3-\alpha_t}{2}& \alpha_4 + \fr{Q-\alpha_2-\alpha_3-\alpha_t}{2} \end{array}\bigg), \nonumber
\end{align}
together with taking
$$
x = - \fr{Q+\alpha_2+\alpha_3+\alpha_t}{2} - \alpha_4 
$$
to get (\ref{6j2}) which proves the identity \rf{6j2} 
in the main part of the text.

\end{document}